\documentclass{article}

\usepackage{arxiv}

\usepackage[utf8]{inputenc} 
\usepackage[T1]{fontenc}    
\usepackage{hyperref}       
\usepackage{url}            
\usepackage{booktabs}       
\usepackage{amsfonts}       
\usepackage{amsmath}
\usepackage{amssymb}
\usepackage{microtype}      
\usepackage{graphicx}
\graphicspath{ {./} }

\title{Electroabsorption by Confined Excitons with Gaussian Interaction Potential}

\author{
 Yuriy D. Sibirmovsky, Ivan S. Vasil'evskii, Nikolay I. Kargin \\
  Nanoengineering in Electronics, Spintronics and Photonics Institute\\
  National Nuclear Research University "MEPhI"\\
  Moscow 115409, Russian Federation \\
  \texttt{ydsibirmovsky@mephi.ru} \\
}

\begin{document}
\maketitle
\begin{abstract}
We consider the effects of electron-hole interaction, 2D confinement and applied electric field on direct allowed transitions in III-V semiconductors, with InGaAs as a study case. Instead of Coulomb interaction, we use Gaussian potential. It is finite at the origin and has a finite effective range, which allows for a more efficient numerical solution of Schr\"{o}dinger equation. Yet, we can expect electroabsorption phenomena to remain qualitatively similar to the ones observed for Coulomb excitons. Moreover, we use variation of parameters to fit both position and magnitude of the first absorption peak to the Coulomb case.

We combine and compare several numerical and approximate methods, including spectral expansion, finite differences, separation of variables and variational approximation.

We find that separation of variables approach works only for quantum well widths smaller than the exciton radius. After separation of variables, finite difference solution of the resulting interaction equation gives a much better agreement with the full spectral solution than naive variational approximation.

We observe that electric field has a critical effect on the magnitudes of exciton absorption peaks, suppressing previously allowed transitions and enhancing forbidden ones. Moreover, for excited states, initially suppressed transitions are enhanced again at higher field strengths.
\end{abstract}

\keywords{Schr\"{o}dinger equation \and absorption \and electroabsorption \and excitons \and Stark effect \and quantum wells \and spectral methods \and finite difference \and variational approximation}

\section{Introduction}
\label{sec1}
Electroabsorption and electrorefraction phenomena play a crucial role in microwave photonics and other areas. To create compact and efficient devices, such as electro-optic modulators, one needs to utilize the properties of semiconductor nanostructures, such as quantum wells (QWs) and superlattices. The research on the influence of electric field on the optical properties of such low dimensional structures had started more than 30 years ago \cite{Miller0, Miller1, Miller2, Klipstein, Kan, Weiner} with the term "Quantum-Confined Stark Effect" being coined by Miller et al. in 1984 \cite{Miller0} to emphasize the wide range of phenomena arising there, compared to the ordinary atomic Stark Effect.

Molecular beam epitaxy allows one to produce a wide range of QW-based heterostructures, combining dozens of layers of varying thickness and chemical composition. For microwave photonics and fiber communication systems one of the most popular QW materials is In$_x$Ga$_{1-x}$As with barriers made of In$_y$Al$_{1-y}$As, while InP is used as substrate. Increasingly often, more complicated InGaAsP compounds are used to achieve independent control of bandgap and lattice constant \cite{Mohseni, Arakawa}, while special attention is paid to the role of asymmetry in increasing efficiency of electroabsoprtion or electrorefraction \cite{Dave, Feng}.

For such complicated multilayered heterostructures, electro-optic effects depend on the layer composition in a non-trivial way, and design of more efficient devices necessarily involves numerical modeling and simulation. While there have been a multitude of studies dedicated to modeling absorption and refraction (with or without applied electric field) of quantum-confined semiconductor structures \cite{Chuang2, Glutsch1, Ahland, Hader}, this is still an on-going research field \cite{Maslov, Sivalertporn, Zolotarev}. The photonics industry requires a combination of different methods: fast algorithms involving various approximations are needed for real time multi-parameter optimization of heterostructures, while slower, but more accurate algorithms should be used to verify the results quantitatively.

However, there are several challenges involved with the task of modeling electroabsorption. The most important effects are electron-hole interaction and quantum confinement. Secondary effects include non-parabolicity, scattering, band mixing or coupling, strain etc. While complex methods which take most of these features into account have already been developed, they are usually quite complicated, require a lot of computational power, and are not straightforward to implement. In fact, the influence of Coulomb interaction even on bulk Franz–Keldysh effect is still a subject of investigation both theoretical \cite{Duque-Gomez} and applied \cite{YueHu}.

Which is why in this paper, we consider a rather basic problem: a single exciton, confined in a quantum well and subjected to electric field in the direction, perpendicular to the QW plane. In addition, we use a modified interaction potential, which, while qualitatively similar to the Coulomb one, is finite at the origin, and has a finite effective range. This leads to a more efficient numerical solution, which in turn allows us to compare different methods and approximations, as well as study the electroabsorption phenomena, and the role played by quantum confinement. This potential can also be useful in modeling the phenomenon of static screening, which becomes important at higher light intensities and/or doping levels. It should be noted that simplified interaction potentials are a common approach to probe complex phenomena, where analytical solutions are not readily available otherwise, see for example \cite{Schmitt} where delta-function potential is used to model exciton interaction.

The main goals of this work are as follows. The first one is to test the widely used separation of variables approach and its particular case, variational approximation for the electron-hole interaction wavefunction. The second is to test the spectral method, which, compared to grid methods (finite differences or finite elements) leads to much smaller and more manageable matrices. The third is to obtain a qualitative picture of electroabsorption in a quantum well, which should be quite similar to the actually observed Coulomb exciton case.

We used R language \cite{Rlanguage} for most of the calculations and for plot generation, while Wolfram Mathematica \cite{Mathematica} was used for the spectral expansion method and for evaluating certain special functions.

We will present all the absorption spectra in absolute units of $\mu$m$^{-1}$, which will allow us to quantitatively compare the magnitude for different cases. The important constants and parameters are summarized in Table 1. We use a two-band parabolic effective mass envelope function approximation in position space. For clarity, in his work we don't take into account polarization dependence of QW absorption and only consider the heavy-hole exciton.

\begin{table}
\caption{Constants and material parameters used in calculation}
\centering
\begin{tabular}{ l c l l }
 \textbf{Name} & \textbf{Parameter} & \textbf{Value} & \textbf{Units} \\ 
 \hline
 - & $\frac{\hbar^2}{m_0}$ & $0.07619964$ & eV $\cdot$ nm$^2$ \\ 
 Fine structure constant & $\alpha_0=\frac{e^2}{4\pi \varepsilon_0 \hbar c}$ & $1/137.036$ & - \\ 
 Bohr radius & $r_0=\frac{\hbar^2}{m_0} \cdot \frac{4\pi \varepsilon_0}{e^2}$ & $0.05291772$ & nm \\ 
 Homogeneous broadening & $\Gamma$ & $0.002$ & eV \\
 \hline
 $ ~ $ & In$_{0.53}$Ga$_{0.47}$As parameters  \cite{Arakawa}: & $ ~ $ & $ ~ $ \\
 \hline
 Bandgap & $E_g$ & $0.721$ & eV \\ 
 Optical matrix parameter & $E_p$ & $23.84$ & eV \\ 
 Backgr. dielectric const. & $\varepsilon_b = n_b^2$ & $13.9$ & - \\ 
 Electron effective mass & $\frac{m_e}{m_0}$ & $0.044$ & - \\ 
 Heavy hole effective mass & $\frac{m_h}{m_0}$ & $0.308$ & - \\ 
 Reduced exciton mass & $\frac{m_{eh}}{m_0}=\frac{m_e m_h}{m_0(m_e+m_h)}$ & $0.0385$ & - \\ 
 Exciton Bohr radius & $r_{ex}=r_0 \frac{\varepsilon_b m_0}{m_{eh}}$ & $19.10536$ & nm \\ 
 Exciton Rydberg energy & $Ry=\frac{\hbar^2}{2m_{eh}r_{ex}^2}$ & $0.002711$ & eV \\ 
 \hline
\end{tabular}
\end{table}

\section{Absorption spectra for bulk and strongly confined excitons}
\label{sec2}

\subsection{Exact expressions}

For bulk (3D) exciton with no applied electric field, the absorption spectrum (due to direct allowed transitions) can be expressed analytically \cite{Tanguy1, Pedersen}:

\begin{align}
\label{eq 1}
\centering
\alpha (\hbar \omega) &= \frac{4}{3} \frac{\alpha_0}{n_b} \frac{\hbar^2}{m_0} \frac{E_p}{r_{ex}} \cdot \Im \left[ \sum_{n=1}^{+\infty} \frac{2/r_{ex}^2 \cdot \hbar w}{E_n \left[E_n^2-(\hbar w)^2 \right] n^3} + \frac{m_{eh}}{\hbar^3 w} G(\hbar w) \right] \nonumber  \\
E_n &= E_g - \frac{Ry}{n^2}  \nonumber  \\
G(\hbar w) &=  g \left(\sqrt{ \frac{Ry}{E_g-\hbar w}}\right)+g \left(\sqrt{ \frac{Ry}{E_g+\hbar w}}\right)-2g \left(\sqrt{ \frac{Ry}{E_g}} \right)  \\
g(\xi) &= \ln \xi + \phi(\xi) + \frac{1}{2 \xi}  \nonumber  \\
\hbar w &= \hbar \omega + i \Gamma \nonumber  
\end{align}

Where $\phi(\xi)$ - digamma function. For a hypothetical 2D exciton confined in a quantum well of with $L$, there's a similar expression \cite{Tanguy2, Pedersen}:

\begin{align}
\label{eq 2}
\centering
\alpha (\hbar \omega) &= \frac{4}{3} \frac{\alpha_0}{n_b} \frac{\hbar^2}{m_0} \frac{E_p}{L} \cdot \Im \left[ \sum_{n=1}^{+\infty} \frac{2/r_{ex}^2 \cdot \hbar w}{\tilde{E}_n \left[\tilde{E}_n^2-(\hbar w)^2 \right] (n-1/2)^3} + \frac{m_{eh}}{\hbar^3 w} F(\hbar w) \right]  \nonumber  \\
\tilde{E}_n &= \tilde{E}_g - \frac{Ry}{(n-1/2)^2}  \nonumber  \\
F(\hbar w) &=  f \left(\sqrt{ \frac{Ry}{\tilde{E}_g-\hbar w}}\right)+f \left(\sqrt{ \frac{Ry}{\tilde{E}_g+\hbar w}}\right)-2f \left(\sqrt{ \frac{Ry}{\tilde{E}_g}} \right)   \\
f(\xi) &= \ln \xi + \phi \left(\xi+\frac{1}{2} \right) \nonumber
\end{align}

Where $\tilde{E}_g$ takes into account the increase of transition energy due to quantum size effect for electrons and holes. 

We may notice the similarity between (\ref{eq 1}) and (\ref{eq 2}), which highlights the fact that strongly confined excitons have ground state binding energy $4$ times greater and peak absorption $8 r_{ex}/L$ times greater than their bulk counterparts.

Using these expressions and taking the real part instead of imaginary, we could also calculate refractive index, with no need to directly apply Kramers-Kronig relations.

If we discard the electron-hole interaction, and further take $\Gamma = 0$, Eq. (\ref{eq 1}) reduces to the well known square root dependence \cite{Chuang1}:

\begin{equation}
\label{eq 3}
\centering
\alpha(\hbar \omega) = \frac{2 \sqrt{2}}{3} \frac{\alpha_0}{n_b} \left( \frac{\hbar^2}{m_0} \right)^{-1/2} \left( \frac{m_{eh}}{m_0} \right)^{3/2} \frac{E_p}{\hbar \omega} \sqrt{\hbar \omega-E_g}
\end{equation}

While (\ref{eq 2}) reduces to the step function $\theta(\xi)$ dependence (in the following expression we only take into account the ground state) \cite{Chuang1}:

\begin{equation}
\label{eq 4}
\centering
\alpha(\hbar \omega) = \frac{2 \pi}{3 L} \frac{\alpha_0}{n_b}  \frac{m_{eh}}{m_0} \frac{E_p}{\hbar \omega} \theta(\hbar \omega-\tilde{E}_g)
\end{equation}

These expressions are important as the way to check the numerical results for modified interaction potentials.

\subsection{Finite difference schemes}

When no confinement is present (3D) or in the case of very strong confinement (2D) and no applied electric field, Schr\"{o}dinger equation reduces to a single variable case, with either spherical or cylindrical symmetry.

\begin{align}
\label{eq 5}
\hat{H} \psi=-\frac{\hbar^2}{2m_{eh}} \Delta_r \psi(r)-2 Ry U(r) \psi(r)=E \psi(r)   \\
U(r)= \left\lbrace \begin{array}{cc}
 \dfrac{r_{ex}}{r} & \text{Coulomb interaction} \\
 \dfrac{1}{a} \exp \left(-\dfrac{r^2}{b^2 r_{ex}^2} \right) & \text{Gaussian interaction}
\end{array} \right. \nonumber
\end{align}

For the Gaussian interaction potential, introduced here, $a$ is the depth parameter and $b$ the width parameter. This combination of parameters allows us to vary both the interaction strength and its range.

Reduction to (effectively) one dimension allows us to apply a finite difference (FD) scheme with small and uniform step size, which is the most simple way to account for any interaction potential. However, in these cases, one needs to be careful when deriving the FD approximations for both kinetic and potential energy operators. In general, the Hamilton operator matrix will not be Hermitian.

Let the grid be defined by $r_j=j \Delta r$ with $j=0,1, \ldots, N_r$ and $N_r \Delta r=R$. We summarize our schemes in Table 2. The 2D scheme mostly follows \cite{Glutsch1}, except for the Coulomb potential approximation. The 3D scheme is derived independently here.

\begin{table}
\caption{Finite difference approximations to operators in 2D and 3D}
\centering
\begin{tabular}{ c c c }
 \textbf{Operator} & \textbf{2D} & \textbf{3D} \\ 
 \hline
 $~$ & $~$ & $~$ \\
  $ \Vert \psi \Vert^2 $ &  $ \int_0^{+\infty} 2 \pi r |\psi(r)|^2 dr$ & $\int_0^{+\infty} 4 \pi r^2 |\psi(r)|^2 dr$ \\ 
   $~$ & $~$ & $~$ \\
 $\Delta_r \psi$ &  $\dfrac{1}{r} \dfrac{\partial}{\partial r} \left(r \dfrac{\partial \psi}{\partial r} \right)$ & $\dfrac{1}{r^2} \dfrac{\partial}{\partial r} \left(r^2 \dfrac{\partial \psi}{\partial r} \right)$ \\
  $~$ & $~$ & $~$ \\
  \hline
  \textbf{FD approx.} & \textbf{2D} & \textbf{3D} \\ 
 \hline
  $~$ & $~$ & $~$ \\
 $\left(\Delta_r \psi \right)_0$ &  $\dfrac{4}{(\Delta r)^2} \left(-\psi_0+\psi_1 \right)$ & $\dfrac{6}{(\Delta r)^2} \left(-\psi_0+\psi_1 \right)$ \\ 
  $~$ & $~$ & $~$ \\
 \hline
  $~$ & $~$ & $~$ \\
  $\left(\Delta_r \psi \right)_{j \geq 1}$ & $~$ & $~$ \\
    2D & \multicolumn{2}{ c }{$\dfrac{1}{(\Delta r)^2} \left[ \left(1-\dfrac{1}{2j} \right)\psi_{j-1}-2\psi_j+\left(1+\dfrac{1}{2j} \right)\psi_{j+1} \right]$ } \\
     $~$ & $~$ & $~$ \\
    3D & \multicolumn{2}{ c }{$\dfrac{1}{(\Delta r)^2} \left[ \left(1-\dfrac{1}{2j} \right)^2\psi_{j-1}-\left(2+\dfrac{1}{2j^2} \right)\psi_j+\left(1+\dfrac{1}{2j} \right)^2\psi_{j+1} \right]$} \\ 
     $~$ & $~$ & $~$ \\
 \hline
  $~$ & $~$ & $~$ \\
 $\left(\dfrac{\psi}{r} \right)_0$ &  $\dfrac{4}{\Delta r} \psi_0$ & $\dfrac{3}{\Delta r} \psi_0$ \\ 
  $~$ & $~$ & $~$ \\
 $\left(\dfrac{\psi}{r} \right)_{j \geq 1}$ &  \multicolumn{2}{ c }{$\dfrac{\psi_j}{j \Delta r} $} \\
  $~$ & $~$ & $~$ \\
 $\left(e^{-\beta^2 r^2} \psi  \right)_{j \geq 0}$ &  \multicolumn{2}{ c }{$e^{-\beta^2 j^2 (\Delta r)^2} \psi_j  $} \\
  $~$ & $~$ & $~$ \\
$ \Vert \psi \Vert^2 $ &  $ 2\pi (\Delta r)^2 \sum_{j=0}^{N_r} s_j |\psi_j|^2 $ & $ 4 \pi (\Delta r)^3 \sum_{j=0}^{N_r} s_j |\psi_j|^2 $ \\ 
 $~$ & $~$ & $~$ \\
$ s_0$ &  $ \dfrac{1}{8} $ & $ \dfrac{1}{24} $ \\ 
 $~$ & $~$ & $~$ \\
$ s_{j \geq 1}$ &  $ j $ & $ j^2+\dfrac{1}{12} $ \\ 
 $~$ & $~$ & $~$ \\
 \hline
\end{tabular}
\end{table}

Let's introduce $\hat{S}$ as a diagonal matrix with elements $s_j$ and $\hat{H}$ as a tridiagonal matrix with elements defined by Table 2. Then equation (\ref{eq 5}) transforms into a generalized matrix eigenvalue problem:

\begin{equation}
\label{eq 6}
\hat{S} ( \hat{H} +E_g)\vec{\psi} = E \hat{S} \vec{\psi}
\end{equation}

It's easy to see that for the schemes in Table 2, the matrix $\hat{S} \hat{H}$ is Hermitian for the 2D exciton, but non-Hermitian in the 3D case. However, the latter asymmetry is not significant, and only requires us to take the real parts of eigenvalues and eigenvectors as a precaution.

After solving (\ref{eq 6}) and obtaining eigenvalues $E_n$ and eigenvectors $\vec{\psi}^{(n)}$ where $n=1,\ldots,N_r+1$, it's convenient to renormalize the eigenvectors in the following way:

\begin{equation}
\label{eq 7}
\vec{\phi} = \frac{\left\{\sqrt{s_0} \psi_0,\sqrt{s_1} \psi_1,\ldots,\sqrt{s_{N_r}} \psi_{N_r} \right\}}{\sqrt{\sum_{j=0}^{N_r} s_j \psi_j^2}}
\end{equation}

Note that the FD approximations at $j=0$ for operators with a singularity at the origin are obtained here by averaging with the appropriate weight function from $0$ to $\Delta r/2$.

The resulting expressions for the absorption spectrum will be:

\begin{equation}
\label{eq 8}
\alpha_{2D} (\hbar \omega) = \frac{4}{3} \frac{\alpha_0}{n_b} \frac{\hbar^2}{m_0} \frac{E_p}{L} \cdot \Im \left[ \sum_{n=1}^{N_r+1} \frac{8/(\Delta r)^2 \cdot \vert \phi_0^{(n)} \vert^2 \cdot \hbar w}{E_n \left[E_n^2-(\hbar w)^2 \right]}\right] 
\end{equation}

\begin{equation}
\label{eq 9}
\alpha_{3D} (\hbar \omega) = \frac{4}{3} \frac{\alpha_0}{n_b} \frac{\hbar^2}{m_0} \frac{E_p}{\Delta r} \cdot \Im \left[ \sum_{n=1}^{N_r+1} \frac{12/(\Delta r)^2 \cdot \vert \phi_0^{(n)} \vert^2 \cdot \hbar w}{E_n \left[E_n^2-(\hbar w)^2 \right]}\right] 
\end{equation}

The continuous part of the spectrum is modeled by the contribution of discrete FD states with $E_n \geq E_g$, provided that $R$ is large enough so that $E_{n+1}-E_n < \Gamma$ and $\Delta r$ is small enough so that $E_{N_r+1} > \hbar \omega_{\text{max}}$, where $\hbar \omega_{\text{max}}$ is the largest photon energy we are considering. Verification of the proposed schemes and formulas (\ref{eq 8}-\ref{eq 9}) is provided in Figure 1(a,b), where the numerical results are compared with the exact expressions (\ref{eq 1}-\ref{eq 2}). For the strongly confined case we take $\tilde{E}_g=E_g$ for convenience.

\begin{figure}
\centering
\includegraphics[width=7.4cm]{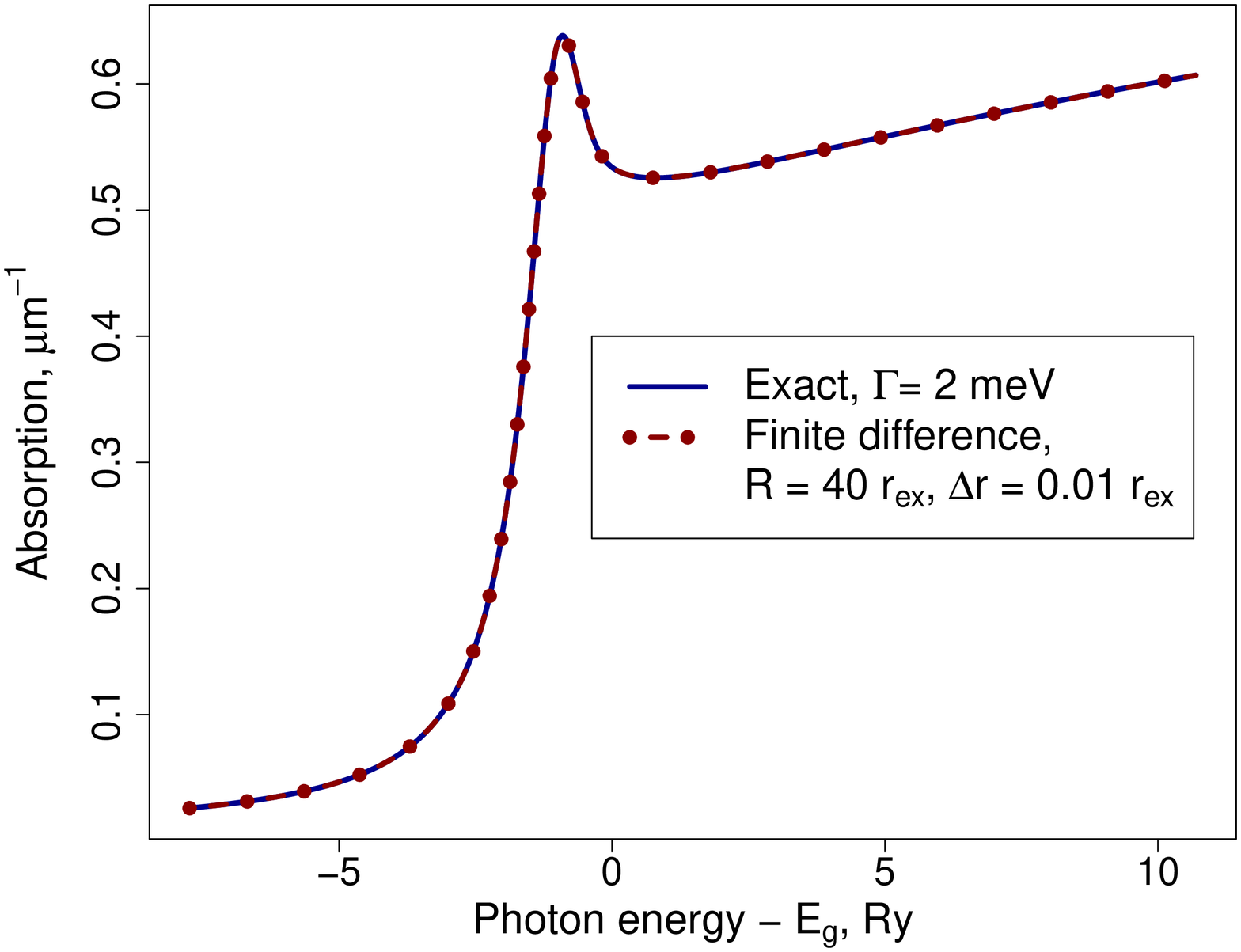}
\includegraphics[width=7.4cm]{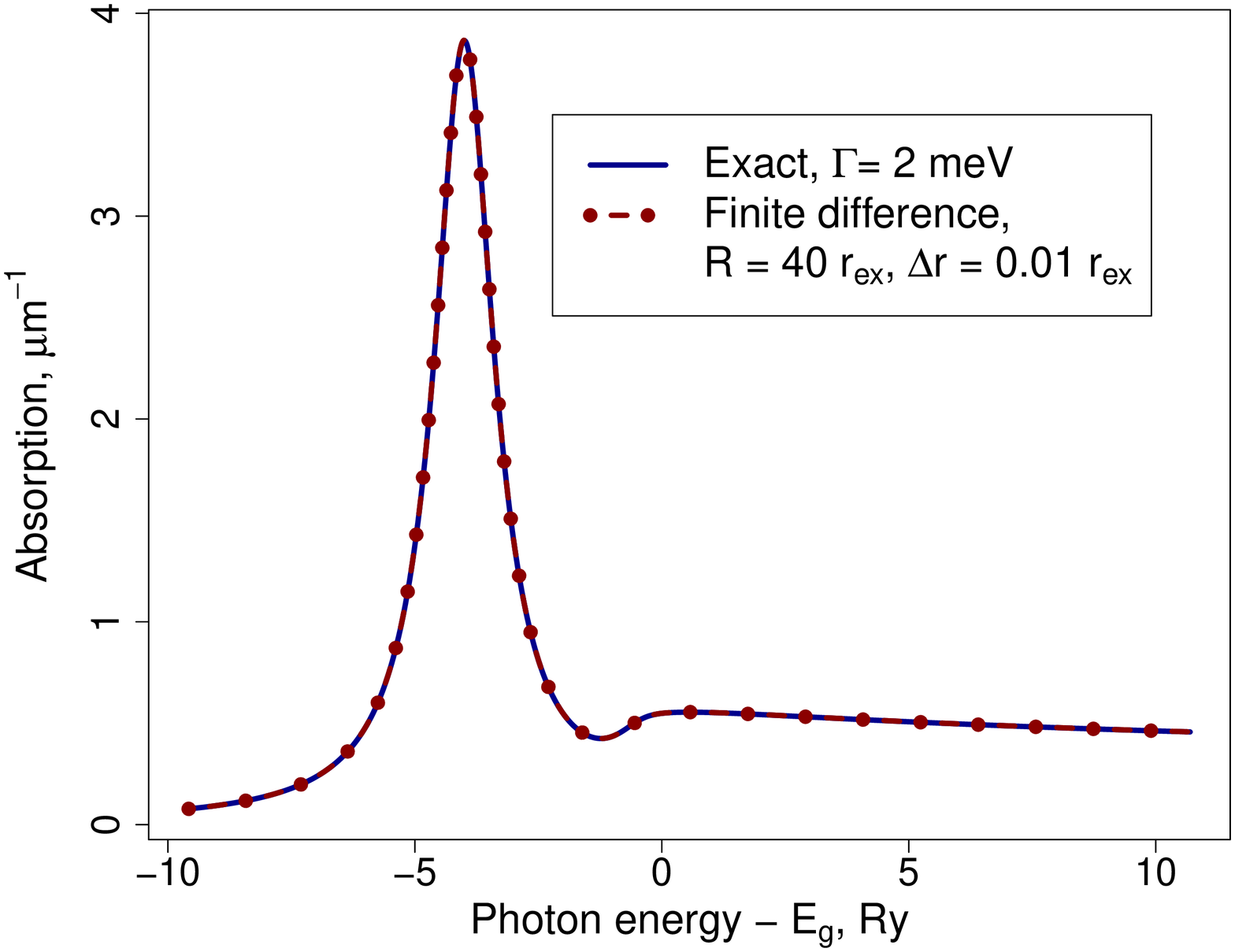}

$ \qquad $ (a)$ \qquad \qquad \qquad \qquad \qquad \qquad \qquad \qquad \qquad  $(b)
\caption{Excitonic absorption spectra for (a) 3D (bulk) and (b) 2D (strongly confined) excitons ($L$ is taken equal to $r_{ex}$), as given by exact formulas (\ref{eq 1}-\ref{eq 2}) and finite difference calculations.}
\end{figure}

\begin{figure}
\centering
\includegraphics[width=7.4cm]{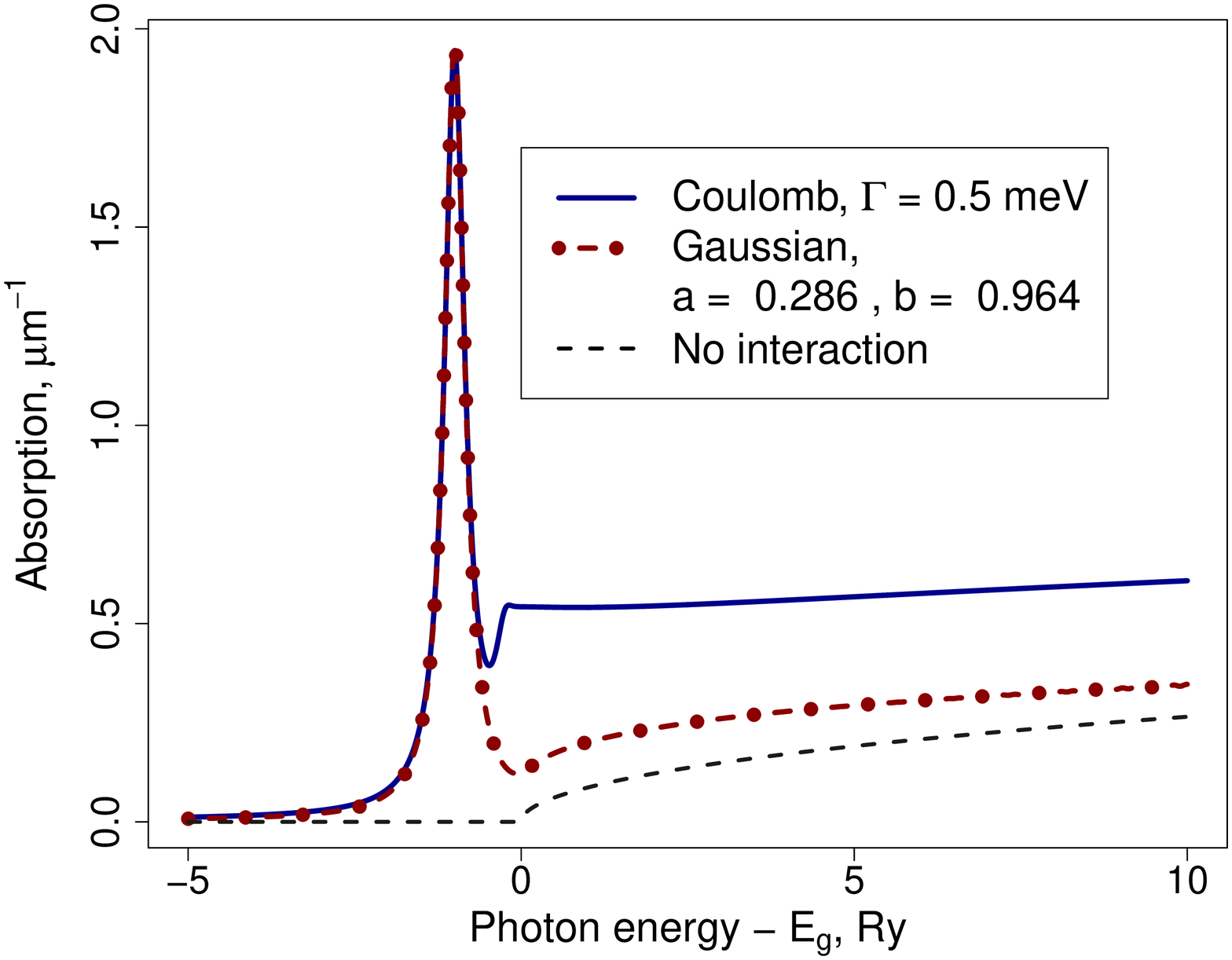}
\includegraphics[width=7.4cm]{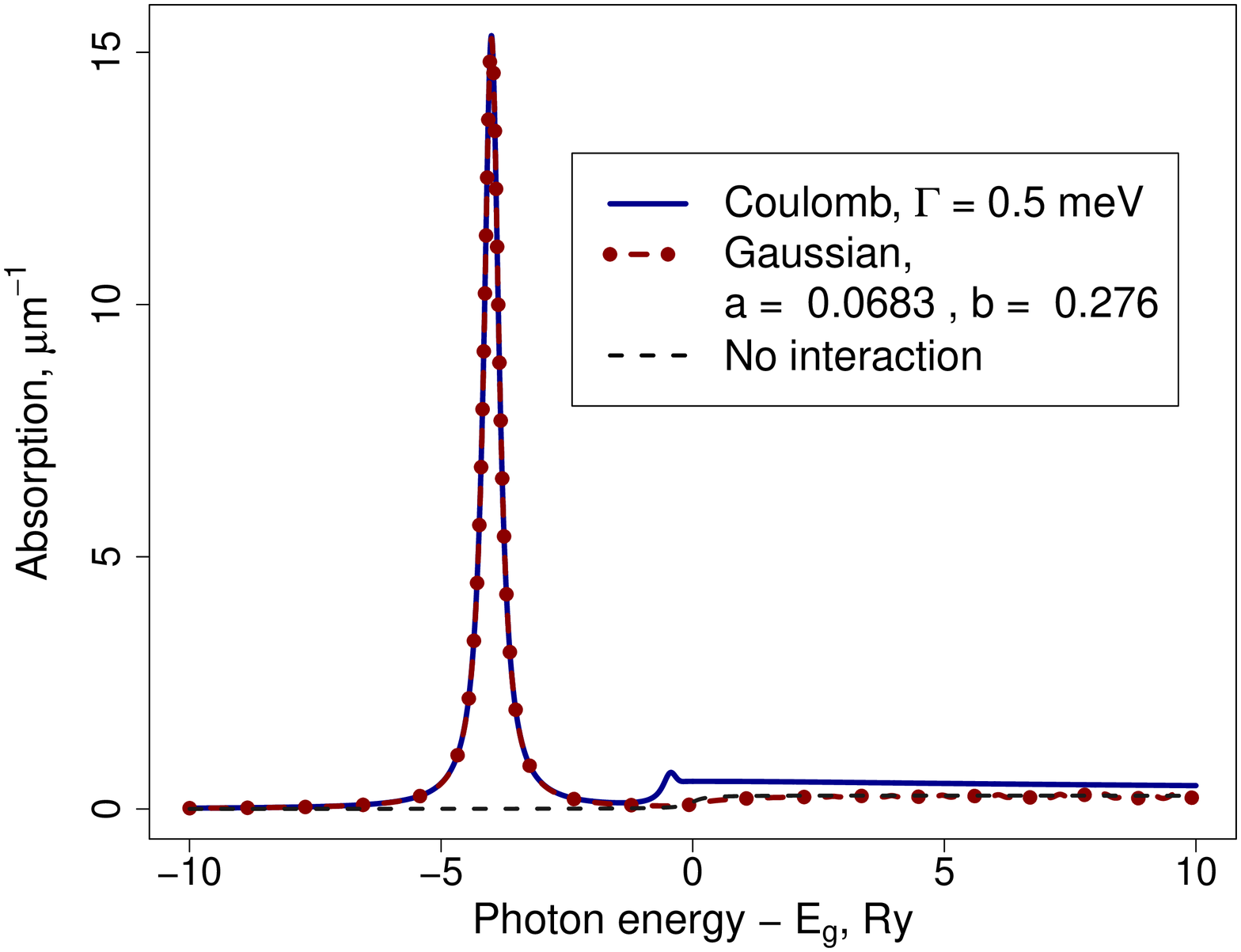}

$ \qquad $ (a)$ \qquad \qquad \qquad \qquad \qquad \qquad \qquad \qquad \qquad  $(b)

\includegraphics[width=7.4cm]{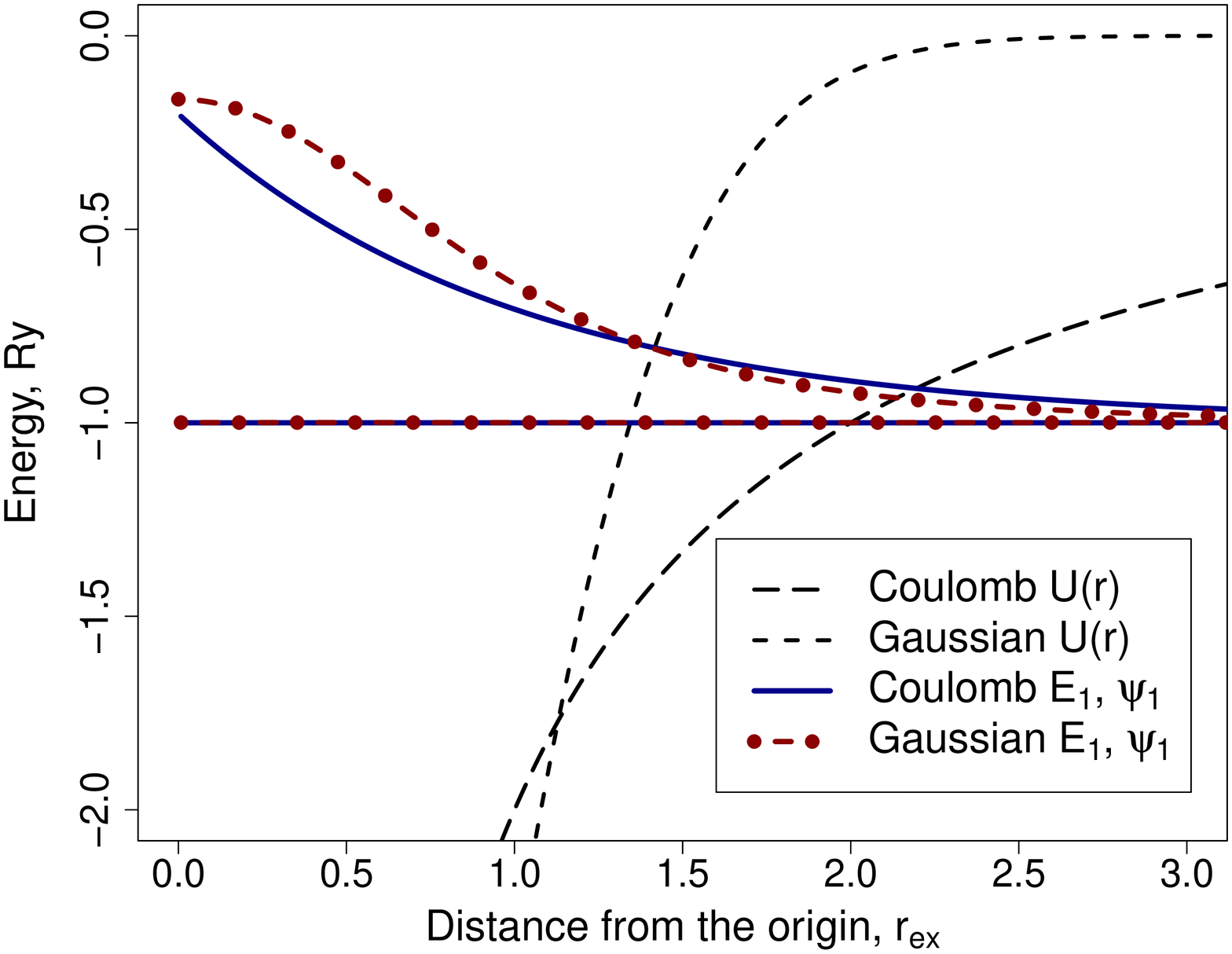}
\includegraphics[width=7.4cm]{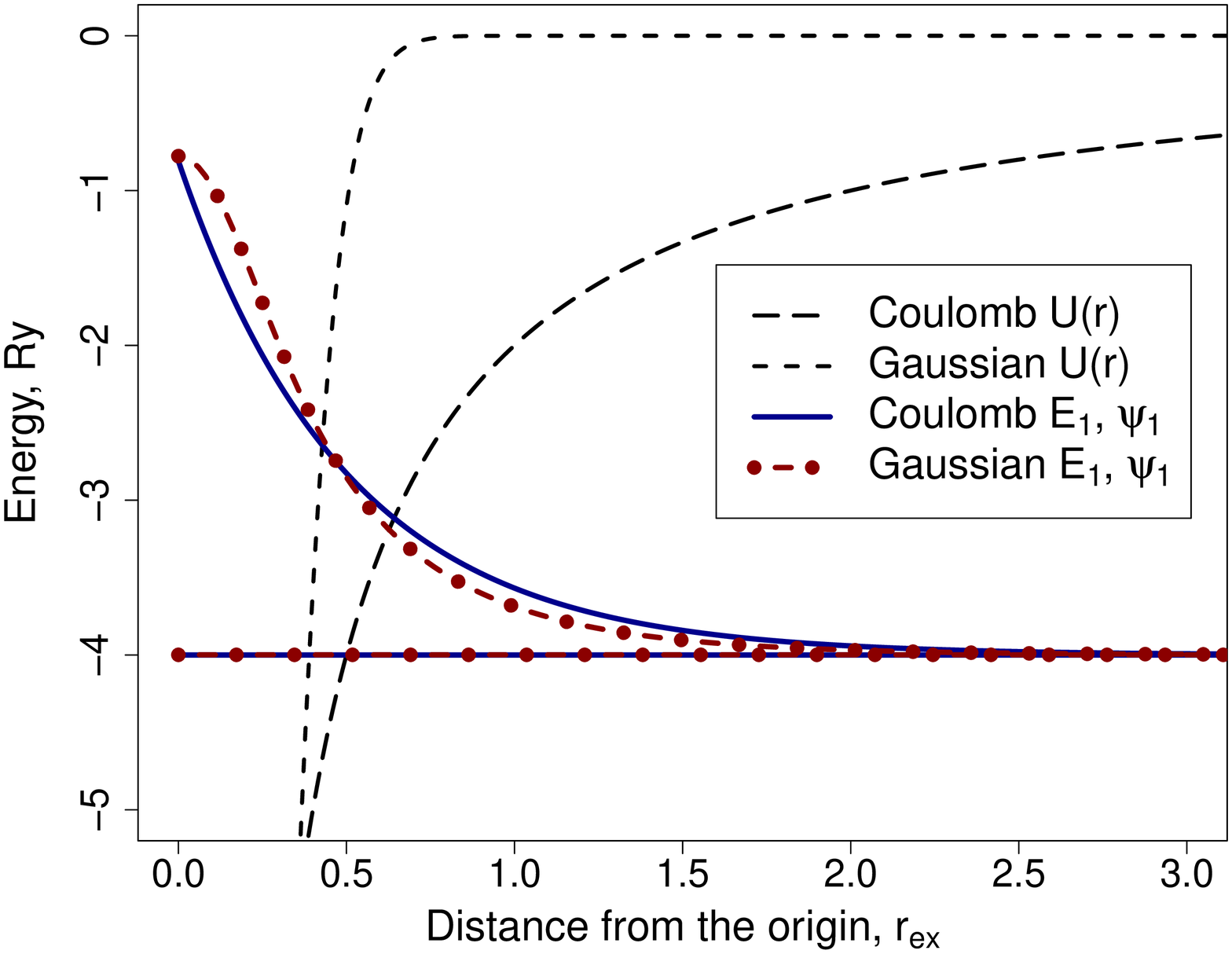}

$ \qquad $ (c)$ \qquad \qquad \qquad \qquad \qquad \qquad \qquad \qquad \qquad  $(d)

\caption{Comparison between Coulomb and Gaussian (with parameters given in Table 3) excitons. Absorption spectra for (a) 3D and (b) 2D excitons. Potentials near the origin and ground state wavefunctions for (c) 3D and (d) 2D excitons.}
\end{figure}

To compare relative magnitudes of the exciton absorption peaks, we introduce a dimensionless parameter, proportional to the transition oscillator strength:

\begin{align}
\label{eq 10}
Q^{2D}_n =\frac{4 r_{ex}^2}{(\Delta r)^2} \cdot \vert \phi_0^{(n)} \vert^2 \\
Q^{3D}_n =\frac{6 r_{ex}^3}{(\Delta r)^3} \cdot \vert \phi_0^{(n)} \vert^2 \nonumber
\end{align}

\subsection{Fitting the parameters of interaction potential}

By varying the depth and width parameters $a,b$ of the Gaussian interaction potential, it is possible to obtain different numbers of bound states, as well as binding energy and amplitude for the ground state. Empirically, we found that a reasonably good agreement with both 3D and 2D Coulomb exciton absorption edges is achieved for the values presented in Table 3. Figures 2(a,b) show the absorption spectra comparison, this time with a small broadening $\Gamma = 0.5$ meV to reduce the contribution from the above-bandgap states, which is present in the 3D case.

In both cases, Gaussian potential supports a single bound state with binding energy and oscillator strength provided in Table 3. The values were calculated with $R = 40 r_{ex}$ and $\Delta r = 0.005 r_{ex}$, and the provided digits are correct.

From Figures 2(c,d) we can see that the wavefunction for the Gaussian potential has zero slope at the origin, which is due to the fact that the potential is finite everywhere. However, for the 2D case, the exponential Coulomb solution agrees very well with the numerical one for the Gaussian potential.

As for the continuous part of the spectrum, due to the short range of the Gaussian potential, there's no significant Sommerfeld enhancement, and this part strongly resembles the no interaction dependence in 3D and (especially) 2D cases.

\begin{table}
\caption{Ground state parameters for 3D and 2D excitons}
\centering
\begin{tabular}{ l c c c c }
 $~$ & a & b & Binding energy $E_1$, Ry & Oscillator strength $Q_1$ \\ 
 \hline
 $~$ & $~$ & $~$ \\
Coulomb 3D & - & - & -1  & 1 \\
Coulomb 2D & - & - & -4  & 8 \\
Gaussian 3D & 0.286 & 0.964 & -0.999  & 1.034 \\
Gaussian 2D & 0.0683 & 0.276 & -3.99  & 7.96 \\
 \hline
\end{tabular}
\end{table}

\section{Confined exciton in electric field. Spectral expansion}
\label{sec3}

Let us consider the problem of interacting electron-hole pair under applied electric field $F$, and confined in the $z$ direction in a QW of width $L$. Then Schr\"{o}dinger equation in relative coordinates will have the form (where we have explicitly added the bandgap energy):

\begin{equation}
\label{eq 11}
-\frac{\hbar^2}{2m_{eh}}\left[\frac{\partial^2 \psi}{\partial z^2}+ \frac{1}{r} \frac{\partial}{\partial r} \left(r \frac{\partial \psi}{\partial r} \right) \right]+eFz \psi-2 Ry U(z,r) \psi+E_g \psi=E \psi
\end{equation}

\begin{equation*}
 -L \leq z \leq L, \qquad 0 \leq r < +\infty 
\end{equation*}

This is a simplified picture, which ignores the contribution of center-of-mass confinement. However, the latter is not affected by electric field and thus is not of interest for the present work. It will be considered in further publications.

Let us expand the wavefunction of equation (\ref{eq 11}), using the following basis:

\begin{align}
\label{eq 12}
\psi(z,r)= \sum_{n_z=1}^{N_z} \sum_{n_r=1}^{N_r} C_{n_z n_r} f_{n_z} (z) g_{n_r}(r) \\
f_{n_z} (z) = \frac{1}{\sqrt{L}} \sin \left[ \frac{\pi n_z}{2} \left(1-\frac{z}{L} \right) \right] \nonumber \\
g_{n_r} (r) = \frac{1}{\sqrt{\pi} R J_1(\gamma_{n_r})} J_0 \left(\gamma_{n_r}\frac{r}{R} \right) \nonumber  
\end{align}

Where $J_0,J_1$ are Bessel functions, $\gamma_{n_r}$ - zeros of $J_0(\xi)$, and we now assume $0 \leq r \leq R$ where $R \gg r_{ex}$. This choice of basis functions allows us to describe both the discrete and continuous parts of the absorption spectrum, similar to the finite difference method. Normalization conditions for the basis functions are:

\begin{equation*}
\int_{-L}^L dz f_{n_z} (z) f_{n'_z} (z) =\delta_{n_z,n'_z} 
\end{equation*}

\begin{equation*}
\int_{0}^R 2\pi r dr g_{n_r} (r) g_{n'_r} (r) =\delta_{n_r,n'_r} 
\end{equation*}

Let us now find the matrix elements of the operators in equation (\ref{eq 11}) in the chosen basis:

\begin{equation}
\label{eq 13}
\left\langle f_{n_z} \left\vert -\frac{\hbar^2}{2m_{eh}} \frac{\partial^2}{\partial z^2} \right\vert f_{n'_z} \right\rangle = \frac{\hbar^2 \pi^2 n_z^2}{8 m_{eh} L^2} \delta_{n_z,n'_z}
\end{equation}

\begin{equation}
\label{eq 14}
\left\langle g_{n_r} \left\vert -\frac{\hbar^2}{2m_{eh}} \frac{1}{r} \frac{\partial}{\partial r} \left(r \frac{\partial}{\partial r} \right) \right\vert g_{n'_r} \right\rangle = \frac{\hbar^2 \gamma_{n_r}^2}{2 m_{eh} R^2} \delta_{n_r,n'_r}
\end{equation}

\begin{equation}
\label{eq 15}
\left\langle f_{n_z} \left\vert eF z \right\vert f_{n'_z} \right\rangle = \frac{8 e F L}{\pi^2} \frac{\left(1-(-1)^{n_z+n'_z} \right)}{(n_z^2-n^{'2}_z)^2} n_z n'_z
\end{equation}

\begin{equation}
\label{eq 16}
\left\langle f_{n_z} \left\vert \exp \left(-\frac{z^2}{b^2 r_{ex}^2} \right) \right\vert f_{n'_z} \right\rangle = \frac{\sqrt{\pi}}{4\nu} \left[V_{n_z-n'_z} \left( \nu \right)-V_{n_z+n'_z} \left( \nu\right) \right]
\end{equation}

\begin{equation*}
\nu=\frac{L}{b r_{ex}} 
\end{equation*}

\begin{equation}
\label{eq 17}
V_n(\nu)=\cos \frac{\pi n}{2} \exp \left(-\frac{\pi^2 n^2}{16 \nu^2}\right) \left[ \text{erf} \left(\nu+i \frac{\pi n}{4 \nu}\right) + \text{erf} \left(\nu-i \frac{\pi n}{4 \nu} \right) \right]
\end{equation}

\begin{equation}
\label{eq 18}
\left\langle g_{n_r} \left\vert \exp \left(-\frac{r^2}{b^2 r_{ex}^2} \right) \right\vert g_{n'_r} \right\rangle = \frac{2}{J_1(\gamma_{n_r})J_1(\gamma_{n'_r})} \int_0^1 t dt e^{-\eta^2 t^2} J_0(\gamma_{n_r} t) J_0(\gamma_{n'_r} t)
\end{equation}

\begin{equation*}
t= \frac{r}{R}, \qquad \eta=\frac{R}{b r_{ex}} 
\end{equation*}

The integral in (\ref{eq 18}) can't be evaluated in analytical form, however, using the condition $R \gg r_{ex}$, we can claim that $\eta \gg 1$, then it follows:

\begin{equation*}
\int_0^1 t dt e^{-\eta^2 t^2} J_0(\gamma_{n_r} t) J_0(\gamma_{n'_r} t) \approx \int_0^{+\infty} t dt e^{-\eta^2 t^2} J_0(\gamma_{n_r} t) J_0(\gamma_{n'_r} t) 
\end{equation*}

From \cite{Integrals} 6.633.2 the latter integral evaluates to (where $I_0$ is modified Bessel function):

\begin{equation*}
\int_0^{+\infty} t dt e^{-\eta^2 t^2} J_0(\gamma_{n_r} t) J_0(\gamma_{n'_r} t) = \frac{1}{2 \eta^2} \exp \left(-\frac{\gamma_{n_r}^2+\gamma_{n'_r}^2 }{4 \eta^2} \right) I_0 \left(\frac{\gamma_{n_r} \gamma_{n'_r}}{2 \eta^2} \right) 
\end{equation*}

Then it follows:

\begin{equation}
\label{eq 19}
\left\langle g_{n_r} \left\vert \exp \left(-\frac{r^2}{b^2 r_{ex}^2} \right) \right\vert g_{n'_r} \right\rangle \approx \frac{1}{ \eta^2 J_1(\gamma_{n_r})J_1(\gamma_{n'_r})} \exp \left(-\frac{\gamma_{n_r}^2+\gamma_{n'_r}^2 }{4 \eta^2} \right) I_0 \left(\frac{\gamma_{n_r} \gamma_{n'_r}}{2 \eta^2} \right)
\end{equation}

Gaussian form of the interaction potential allows us to express its matrix element as a product of two previously found matrix elements:

\begin{equation}
\label{eq 20}
\left\langle f_{n_z} g_{n_r} \left\vert U(z,r) \right\vert g_{n'_r} f_{n'_z} \right\rangle = \frac{1}{a} \left\langle f_{n_z} \left\vert \exp \left(-\frac{z^2}{b^2 r_{ex}^2} \right) \right\vert f_{n'_z} \right\rangle \left\langle g_{n_r} \left\vert \exp \left(-\frac{r^2}{b^2 r_{ex}^2} \right) \right\vert g_{n'_r} \right\rangle
\end{equation}

Collecting all the matrix elements, we again obtain a matrix eigenvalue problem, with matrix dimensions being $N_z N_r \times N_z N_r$. To be able to evaluate special functions with good accuracy and precision, we used Wolfram Mathematica for this calculation. After solving the eigenvalue problem, we obtain the set of eigenvectors and eigenvalues:

\begin{equation*}
\left\{ E_{\lambda}, \vec{\psi}_{\lambda} \right\}, \qquad \lambda=1,2,\ldots,N_z N_r 
\end{equation*}

Then the absorption spectrum can be expressed as:

\begin{equation}
\label{eq 21}
\alpha (\hbar \omega) = \frac{4}{3} \frac{\alpha_0}{n_b} \frac{\hbar^2}{m_0} \frac{E_p}{L} \cdot \Im \left[ \sum_{\lambda=1}^{N_z N_r} \frac{2/r_{ex}^2 \cdot Q_{\lambda} \cdot \hbar w}{E_{\lambda} \left[E_{\lambda}^2-(\hbar w)^2 \right]}\right]
\end{equation}

Where we again introduce a dimensionless oscillator strength, which now has the form:

\begin{align}
\label{eq 22}
Q^{2D}_{\lambda}= \frac{r_{ex}^2}{R^2} \left( \sum_{n_z=1}^{N_z} \sum_{n_r=1}^{N_r} C^{(\lambda)}_{n_z n_r} \frac{1}{J_1(\gamma_{n_r})} \sin \frac{\pi n_z}{2}  \right)^2  \\
Q^{3D}_{\lambda}= \frac{r_{ex}}{L} Q^{(\lambda)}_{2D} \nonumber
\end{align}

Note that nothing prevents us from making the parameters $a$ and $b$ functions of $L$ to fit the Coulomb absorption edge. However, lacking the full numerical solution for the latter case, we will use the separation of variables approach for the fit, which will be presented in the following section.

\section{Separation of variables approach. Variational approximation}
\label{sec4}

\subsection{Gaussian interaction potential}

Let's assume that variables in equation (\ref{eq 11}) can be separated. This could be achieved by using the average of interaction potential over the $z$ variable. Then we have:

\begin{align}
\label{eq 23}
\psi(z,r)=\xi(z) \zeta (r) \nonumber \\
-\frac{\hbar^2}{2 m_{eh}} \frac{\partial^2 \xi}{\partial z^2}+eF z \xi+E_g \xi=E_z \xi    \\
-\frac{\hbar^2}{2 m_{eh}} \frac{1}{r} \frac{\partial}{\partial r} \left(r \frac{\partial \zeta}{\partial r} \right)-Ry U_0[\xi] \exp \left(-\frac{r^2}{b^2 r_{ex}^2} \right) \zeta=E_r \zeta  \nonumber   \\
U_{0}[\xi] = \frac{2}{a} \int_{-L}^L  \exp \left(-\frac{z^2}{b^2 r_{ex}^2}  \right)  \vert \xi(z) \vert^2 dz  \nonumber 
\end{align}

We can use the same basis for $\xi(z)$ as in (\ref{eq 12}). When there's no electric field, this basis is itself the solution to the first equation in (\ref{eq 23}). Under electric field, the coefficients are found as usual by numerical matrix diagonalization. 

Now let us find the average of interaction potential $U_0[\xi]$ explicitly:

\begin{equation}
\label{eq 24}
U_{0}[\xi]= \frac{\sqrt{\pi}}{2a \nu} \sum_{n_z,n'_z=1}^{N_z} C_{n_z} C_{n'_z} \left[V_{n_z-n'_z} \left( \nu \right)-V_{n_z+n'_z} \left( \nu \right) \right]
\end{equation}

We will use two possible ways to solve the second equation in (\ref{eq 23}), which can be called the electron-hole interaction equation. The first is finite difference method, as presented in 2.3. It's straightforward to implement in this case. The absorption spectrum then has the form:

\begin{equation}
\label{eq 25}
\alpha (\hbar \omega) = \frac{4}{3} \frac{\alpha_0}{n_b} \frac{\hbar^2}{m_0} \frac{E_p}{L} \cdot \Im \left[ \sum_{\lambda_z=1}^{N_z} \sum_{\lambda_r=1}^{N_r+1} \frac{2/r_{ex}^2 \cdot Q^{2D}_{\lambda_z \lambda_r} \cdot \hbar w}{(E_{\lambda_z} +E_{\lambda_r})\left[(E_{\lambda_z} +E_{\lambda_r})^2-(\hbar w)^2 \right]}\right]
\end{equation}

Where:

\begin{equation}
\label{eq 26}
Q^{2D}_{\lambda_z \lambda_r}= \frac{4 r_{ex}^2}{(\Delta r)^2}  \vert \phi_0^{(\lambda_z \lambda_r)} \vert^2
\end{equation}

We need to numerically solve the interaction equation for each state $\lambda_z$.

\subsection{Variational approximations}

The second way to solve (\ref{eq 23}) is variational method, which was one of the first employed for the case of Coulomb excitons \cite{Miller1, Ishikawa} and is still being used to this day \cite{Zolotarev}. In the present case, the most simple choice of ansatz for the ground state is a Gaussian function (this choice is also supported by the shape of numerically found ground state wavefunction in Figure 2(c)).

\begin{align}
\label{eq 27}
\zeta_1(r)=\sqrt{\frac{2}{\pi}} c e^{-c^2 r^2}  \\
\int_0^{+\infty} 2 \pi r dr ~ |\zeta_1(r)|^2=1 \nonumber
\end{align}

Let's find the energy in this state:

\begin{equation}
\label{eq 28}
E_1(c)= \left\langle \zeta_1 \left\vert H_r \right\vert \zeta_1 \right\rangle = \frac{\hbar^2 c^2}{m_{eh}}-Ry U_0 \frac{2 b^2 r_{ex}^2 c^2}{1+2 b^2 r_{ex}^2 c^2}
\end{equation}

Minimization of $E_1(c)$ leads to a simple quadratic equation with exact solution:

\begin{equation}
\label{eq 29}
c=\frac{1}{\sqrt{2} b r_{ex}} \sqrt{b\sqrt{U_0}-1}
\end{equation}

\begin{equation}
\label{eq 30}
E_1(c)=-\frac{Ry}{b^2} \left(b\sqrt{U_0}-1 \right)^2
\end{equation}

Then the variational absorption spectrum will have the following form:

\begin{equation}
\label{eq 31}
\alpha (\hbar \omega) = \frac{4}{3} \frac{\alpha_0}{n_b} \frac{\hbar^2}{m_0} \frac{E_p}{L} \cdot \Im \left[ \sum_{\lambda_z=1}^{N_z} \frac{2/r_{ex}^2 \cdot Q_{1 \lambda_z} \cdot \hbar w}{(E_{\lambda_z} +E_{1 \lambda_z})\left[(E_{\lambda_z} +E_{1 \lambda_z})^2-(\hbar w)^2 \right]}\right]
\end{equation}

Where:

\begin{equation}
\label{eq 32}
Q_{1 \lambda_z}= \frac{1}{b^2} \left(b\sqrt{U_0[\xi_{\lambda_z}]}-1 \right) \left( \sum_{n_z=1}^{N_z}  C^{(\lambda_z)}_{n_z} \sin \frac{\pi n_z}{2}  \right)^2
\end{equation}

When $F=0$ the expressions simplify to:

\begin{align}
\label{eq 33}
U_0[\xi_{\lambda_z}]= \frac{\sqrt{\pi}}{2 a \nu} \left[V_0 \left( \nu\right)-V_{2 \lambda_z} \left( \nu \right) \right] \nonumber \\
Q_{1 \lambda_z}= \frac{1}{b^2} \left(b\sqrt{U_0[\xi_{\lambda_z}]}-1 \right) \\
E_{\lambda_z} +E_{1 \lambda_z}=E_g+\frac{\hbar^2 \pi^2 \lambda_z^2}{8 m_{eh} L^2}-\frac{Ry}{b^2} \left(b\sqrt{U_0[\xi_{\lambda_z}]}-1 \right)^2 \nonumber 
\end{align}

Note that for $\lambda_z=1$ we have the following relations:

\begin{align}
\label{eq 34}
b = \sqrt{-\frac{E_1}{Ry Q_1^2}} \\
a = \frac{b^2}{(b^2 Q_1+1)^2} \frac{\sqrt{\pi}}{2 \nu} \left[V_0 \left( \nu\right)-V_{2} \left( \nu \right) \right] \nonumber
\end{align}

Which allow us to fit the interaction potential parameters to any fixed values of $E_1,Q_1$.

This ansatz, however, doesn't accurately describe the strong confinement case, as can be seen from Figure 2(d), where the ground state wavefunction doesn't have Gaussian shape. Thus, we will use a more advanced choice:

\begin{equation}
\label{eq 35}
\zeta_1(r)=\sqrt{\frac{2}{\pi}} \left( \frac{1}{p^2}+\frac{4 s}{p^2+q^2}+\frac{s^2}{q^2} \right)^{-1/2} \left(e^{-p^2 r^2}+s e^{-q^2 r^2} \right)
\end{equation}

\begin{multline}
\label{eq 36}
E_1(p,q,s)=2 Ry r_{ex}^2 \left( \frac{1}{p^2}+\frac{4 s}{p^2+q^2}+\frac{s^2}{q^2} \right)^{-1} \cdot \\ 
\cdot \left[1+s^2+\frac{8 s p^2 q^2}{(p^2+q^2)^2}-b^2 U_0 \left( \frac{1}{2b^2 r_{ex}^2 p^2+1}+\frac{2s}{b^2 r_{ex}^2 (p^2+q^2)+1}+\frac{s^2}{2b^2 r_{ex}^2 q^2 +1} \right) \right] 
\end{multline}

Instead of explicitly calculating partial derivatives of $E_1(p,q,s)$ and solving the resulting system of non-linear algebraic equations, we will use the numerical minimization routine implemented in Wolfram Mathematica.

\begin{equation}
\label{eq 37}
Q_1(p,q,s)=2 r_{ex}^2 (1+s)^2 \left( \frac{1}{p^2}+\frac{4 s}{p^2+q^2}+\frac{s^2}{q^2} \right)^{-1} 
\end{equation}

In Section 5 we will denote the results obtained from ansatz (\ref{eq 27}) as "variational 1" and from (\ref{eq 35}) as "variational 2".

\subsection{Coulomb interaction potential}

In this case instead of the last two equations in (\ref{eq 23}) we have the following:

\begin{align}
\label{eq 38}
-\frac{\hbar^2}{2 m_{eh}} \frac{1}{r} \frac{\partial}{\partial r} \left(r \frac{\partial \zeta}{\partial r} \right)-Ry U[\xi](r) \zeta=E_r \zeta  \\
U[\xi](r) = 2 r_{ex} \int_{-L}^L  \frac{\vert \xi(z) \vert^2}{\sqrt{r^2+z^2}} dz  \nonumber
\end{align}

In this case the exact explicit expression for the averaged potential in the chosen basis is not straightforward to derive, however there's a way to avoid oscillatory integrals, which we will present here.

\begin{equation}
\label{eq 39}
U[\xi](r) = 2 \frac{r_{ex}}{L} \sum_{n_z,n'_z=1}^{N_z} C_{n_z} C_{n'_z} \left[W_{n_z-n'_z} \left( \mu \right)-W_{n_z+n'_z} \left( \mu  \right) \right]
\end{equation}

Where $\mu=r/L$ and:

\begin{equation}
\label{eq 40}
W_m \left( \mu  \right) = \cos \frac{\pi m}{2} \int_0^1 \frac{\cos \left( \frac{\pi m}{2} t \right)}{\sqrt{t^2+\mu^2}} dt
\end{equation}

For odd $m$ it follows that $W_m (\mu) = 0$, and for even $m$ we have:

\begin{equation}
\label{eq 41}
W_m \left( \mu  \right) = K_0 \left(\frac{\pi |m|}{2} \mu \right)-\cos \frac{\pi m}{2} \int_0^{+\infty} \frac{u J_0 (\mu u) e^{-u}}{u^2+\pi^2 m^2/4} du
\end{equation}

For the special case $m=0$:

\begin{equation}
\label{eq 42}
W_0 \left( \mu  \right) = \text{asinh} \frac{1}{\mu}
\end{equation}

For finite difference approximation at $j=0$ we again use the averaging with weight function $2\pi r$ from $r=0$ to $r=\Delta r/2$, which leads to the following expressions:

\begin{multline}
\label{eq 43}
\left[ W_m \left( \frac{r}{L}  \right) \right]_{j=0} = \frac{4 L}{\Delta r} \cdot \\ \cdot \left[ \frac{1}{\pi |m|} \left( \frac{2 L}{\pi |m| \Delta r}- K_1 \left(\frac{\pi |m| \Delta r}{2 L} \right)  \right)-\cos \frac{\pi m}{2} \int_0^{+\infty} \frac{J_1 (\frac{\Delta r}{2L} u) e^{-u}}{u^2+\pi^2 m^2/4} du \right] 
\end{multline}

\begin{equation}
\label{eq 44}
\left[ W_0 \left( \frac{r}{L}  \right) \right]_{j=0} = \text{asinh} \left( \frac{2L}{\Delta r} \right) +\frac{1}{1+\sqrt{1+(\Delta r)^2/(4 L^2)}}
\end{equation}

Where $J_0,J_1,K_0,K_1$ are Bessel functions and modified Bessel functions. Integration w.r.t. $u$ was performed using Gauss-Laguerre quadrature, which allowed to achieve the accuracy of at least $5$ digits using $35$ quadrature points.

We did not attempt variational approximation for this case, as it offers no advantage over finite difference method, due to the complexity of integrals involved for any choice of variational ansatz.

\subsection{Fitting the parameters of interaction potential}

The results of previous sections were used to fit the parameters $a,b$ of the Gaussian potentials to the Coulomb case, but only for $L < r_{ex}$, as previous calculations showed that separation of variables approach only works in this range.

To fit the parameters, we first calculated the $L$ dependence of the Coulomb absorption spectrum, as described in the previous section. In particular, we were interested in the ground state binding energy $E_1$ and oscillator strength parameter $Q_1$.  Then we used (\ref{eq 34}) to obtain the initial fit of the parameters $a,b$. However, since the first variational ansatz doesn't give very accurate results, we corrected them for a better fit, comparing them with the pure 2D parameters from Table 3. Finally, we utilized the second variational ansatz to check the results and further correct the parameters.

Our goal is to obtain a good agreement between the energies and oscillator strengths (absorption peak magnitudes) in the whole range of quantum well widths, but fitting them using approximate solution for the Coulomb potential leads to incorrect results for $L>r_{ex}$. In the end, we had to use a piecewise function for $a(L/r_{ex})$ and $b(L/r_{ex})$ which is described below:

\begin{equation}
\label{eq 45}
a(y)= \left\lbrace \begin{array}{cc}
a_{2D}+ \sum_{m=1}^3 A_m y^m/(v^m +y^m), & y \leq 3/5 \\ A_4 y^2+A_5 y+A_6, & 3/5 < y \leq 1 \\ a_{3D}-A_7/(v^3+y^3), & y>1.
\end{array} \right.
\end{equation}

\begin{equation}
\label{eq 46}
b(y)= \left\lbrace \begin{array}{cc}
b_{2D}+ \sum_{m=1}^3 B_m y^m/(v^m +y^m), & y \leq 3/5 \\ B_4 y^2+B_5 y+B_6, & 3/5 < y \leq 1 \\ b_{3D}-B_7/(v^3+y^3), & y>1.
\end{array} \right.
\end{equation}

Where $a_{2D}, a_{3D}, b_{2D}, b_{3D}$ are taken from Table 3, $y=L/r_{ex}$, $v=3/2$, the parameters $A_{1-3},B_{1-3}$ were obtained by least squares fit to the Coulomb results for $E_1,Q_1$ at $eF=0$, while $A_{4-7},B_{4-7}$ were used to obtain continuous and monotonic functions, increasing from $a_{2D} (b_{2D})$ at $L=0$ to $a_{3D} (b_{3D})$ at $L \to \infty$. The values for all the parameters are provided in Table 4.

\begin{table}
\caption{Parameters for (\ref{eq 45}) and (\ref{eq 46})}
\centering
\begin{tabular}{ l r | l r c }

$A_1$ & $0.64268$ & $B_1$ & $1.9963$ & $~$ \\
$A_2$ & $-0.31437$ & $B_2$ & $-1.9531$ & $~$ \\
$A_3$ & $0.42149$ & $B_3$ & $2.5899$ & $~$ \\
$A_4$ & $-0.25929$ & $B_4$ & $-0.71791$ & $~$ \\
$A_5$ & $0.52428$ & $B_5$ & $1.4984$ & $~$ \\
$A_6$ & $0.012688$ & $B_6$ & $0.092137$ & $~$ \\
$A_7$ & $0.036395$ & $B_7$ & $0.39957$ & $~$ \\
\end{tabular}
\end{table}

\section{Calculation results and discussion}
\label{sec5}

Note that when presenting the calculation results, we subtract the following expression from the energy (which in the figures' labels we denote simply by $E_g$):

\begin{equation*}
\tilde{E}_g=E_g+\frac{\hbar^2 \pi^2}{8 m_{eh} L^2} 
\end{equation*}

This allows us to discard the strong $L$ dependence of the QW ground state and focus instead on the binding energy change, when comparing the results for QWs of different widths.

The implementation details are as follows. All the FD calculations were performed using R language (base package only) on an Intel ® Core™ i5-3210M CPU @ 2.50GHz $\times$ 4 laptop with 8 Gb RAM. In this case the most time consuming task was numerical matrix diagonalization, though for Coulomb matrix elements, their evaluation required a lot of time as well, due to numerical integration and special functions involved. The duration of full spectrum calculation when using separation of variables approach depends on the number of states accounted for in the $z$ direction, as we have to solve the interaction equation for each of them. Overall, depending on the above conditions, and for grid sizes of $1000-2000$ the whole spectrum evaluation procedure took from around $2$ to $20$ minutes.

As for the spectral method, the calculations were performed using Intel® CoreTM i7-4930MX CPU @ 3.00GHz laptop with 32 Gb RAM. In this case the most time consuming task was evaluation of matrix elements. We have utilized the symmetry, which required us to only evaluate the main diagonal and the upper triangular part of the Hamiltonian matrix, which for two variables is equivalent to the condition $N_r (n_z-1)+n_r \geq N_r (n_z'-1)+n_r'$. Wolfram Mathematica's \textit{Parallelize} function was used to speed up the process. When computing the spectrum itself, the summation was restricted only to the states with the energy between $\hbar \omega_{min}$ and $\hbar \omega_{max}$. For $N_z=30$ and $N_r=150$ the whole procedure took about an hour. For $N_z=50$ and $N_r=150$ approximately $2.5$ hours, and for $N_z=50$ and $N_r=180$ up to $4-5$ hours.

However, it should be noted that computation of spectra for different values of electric field doesn't require evaluating the Hamiltonian matrix more than once, as the electric field strength is just a scalar factor in the equation. Which is why in this case, after initial matrix evaluation had been completed, each spectrum took less than $10$ minutes to compute, making this method very efficient for modeling electro-optic phenomena.

\subsection{Dependence on QW width}

\begin{figure}
\centering
\includegraphics[width=7.4cm]{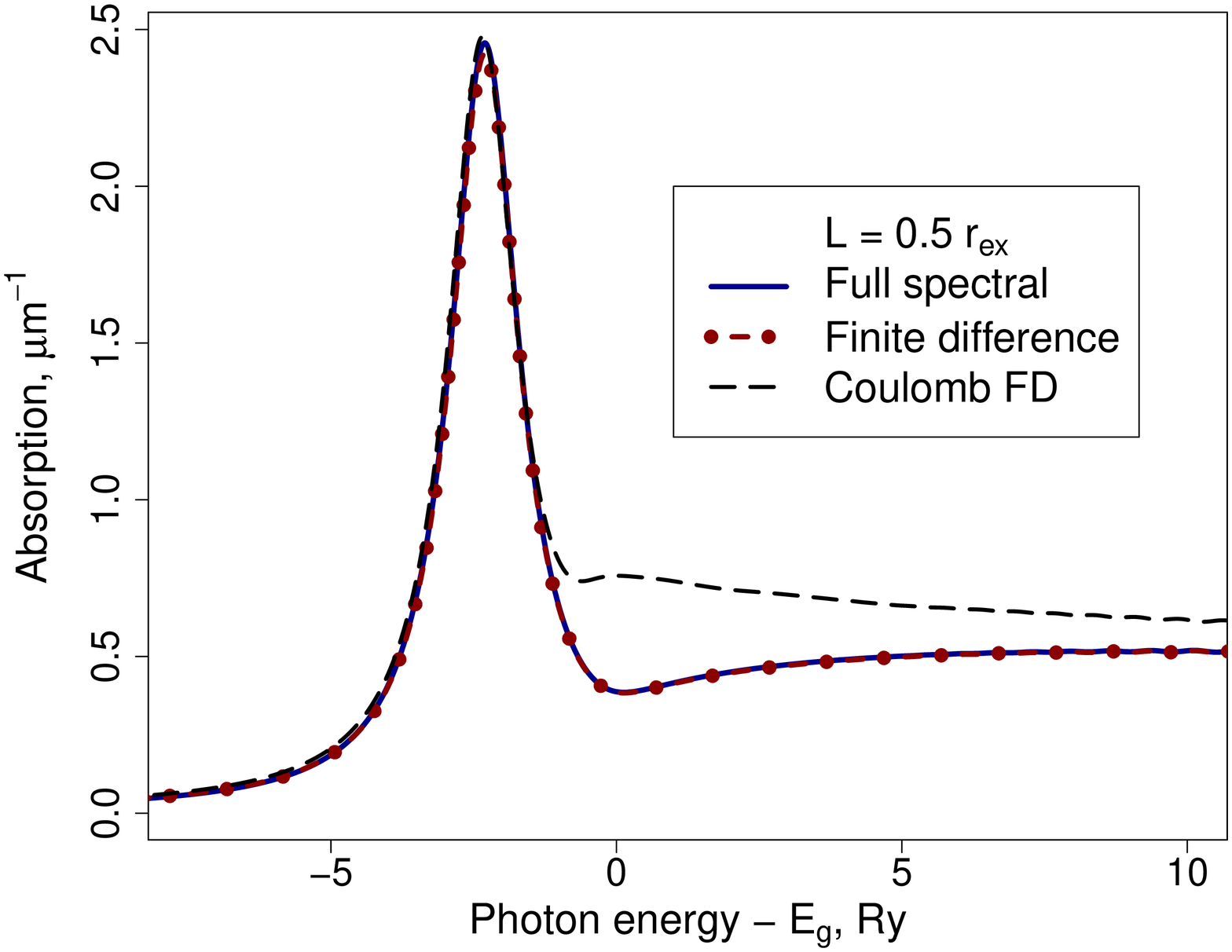}
\includegraphics[width=7.4cm]{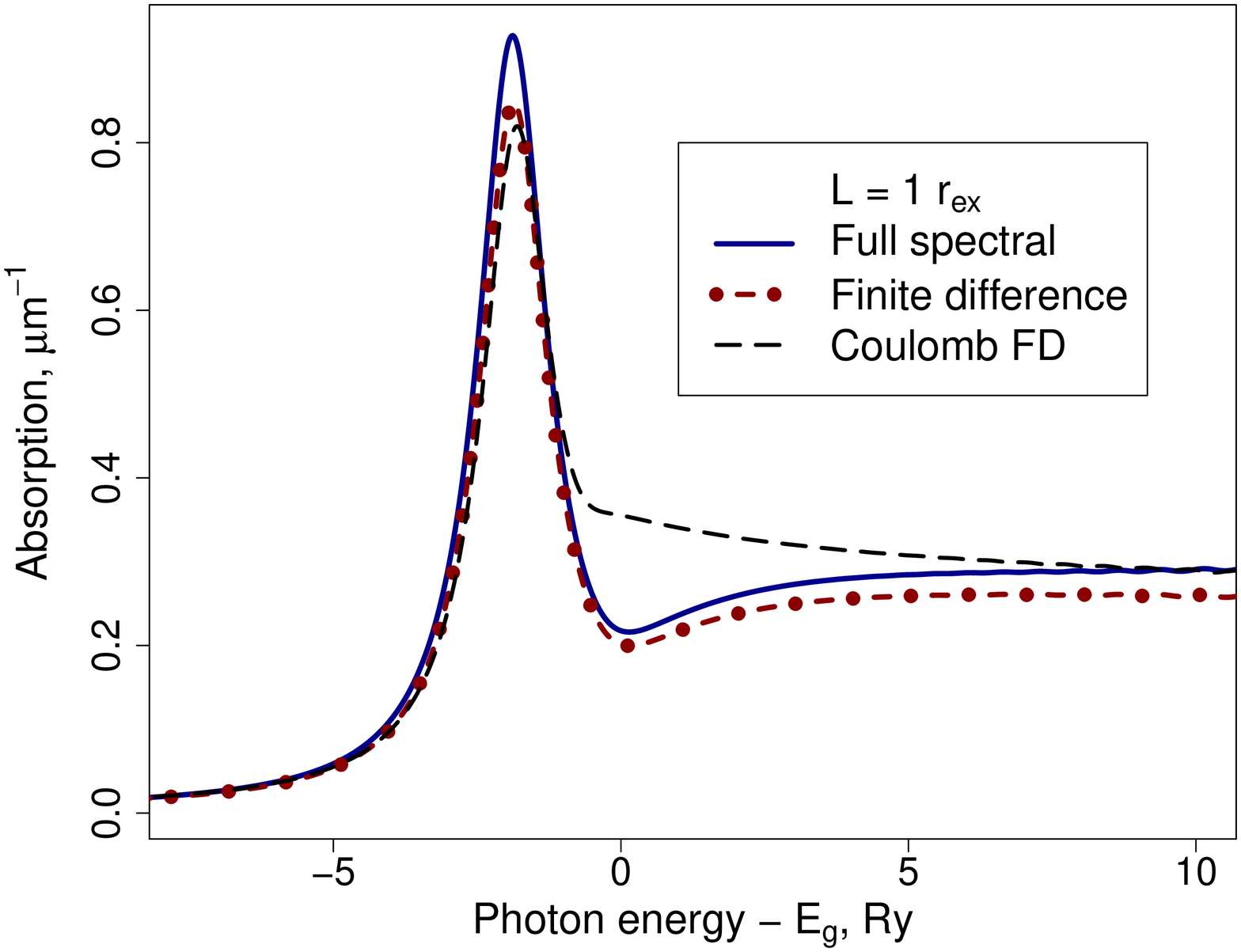}
$ \qquad $  (a)$ \qquad \qquad \qquad \qquad \qquad \qquad \qquad \qquad \qquad  $(b)

\includegraphics[width=7.4cm]{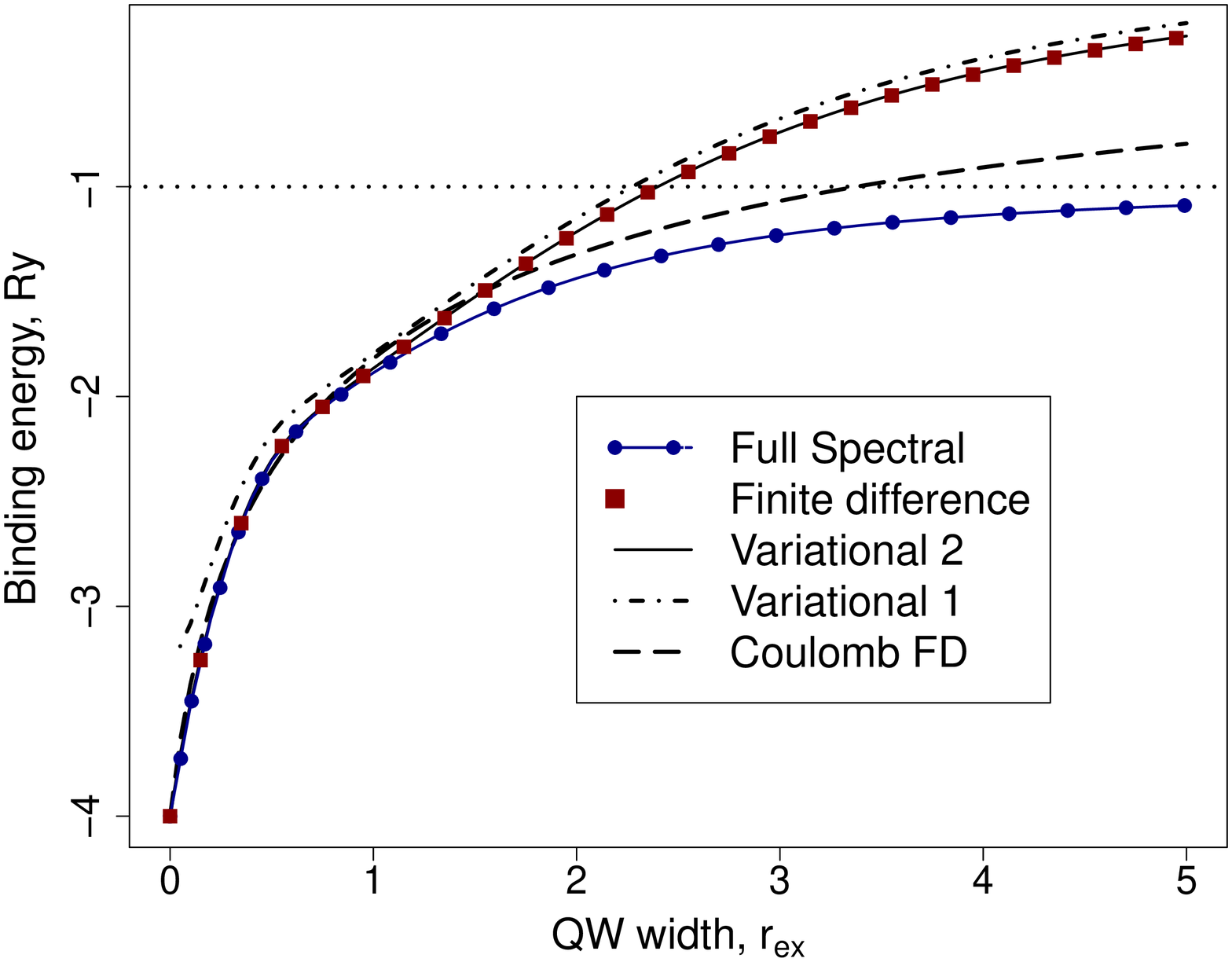}
\includegraphics[width=7.4cm]{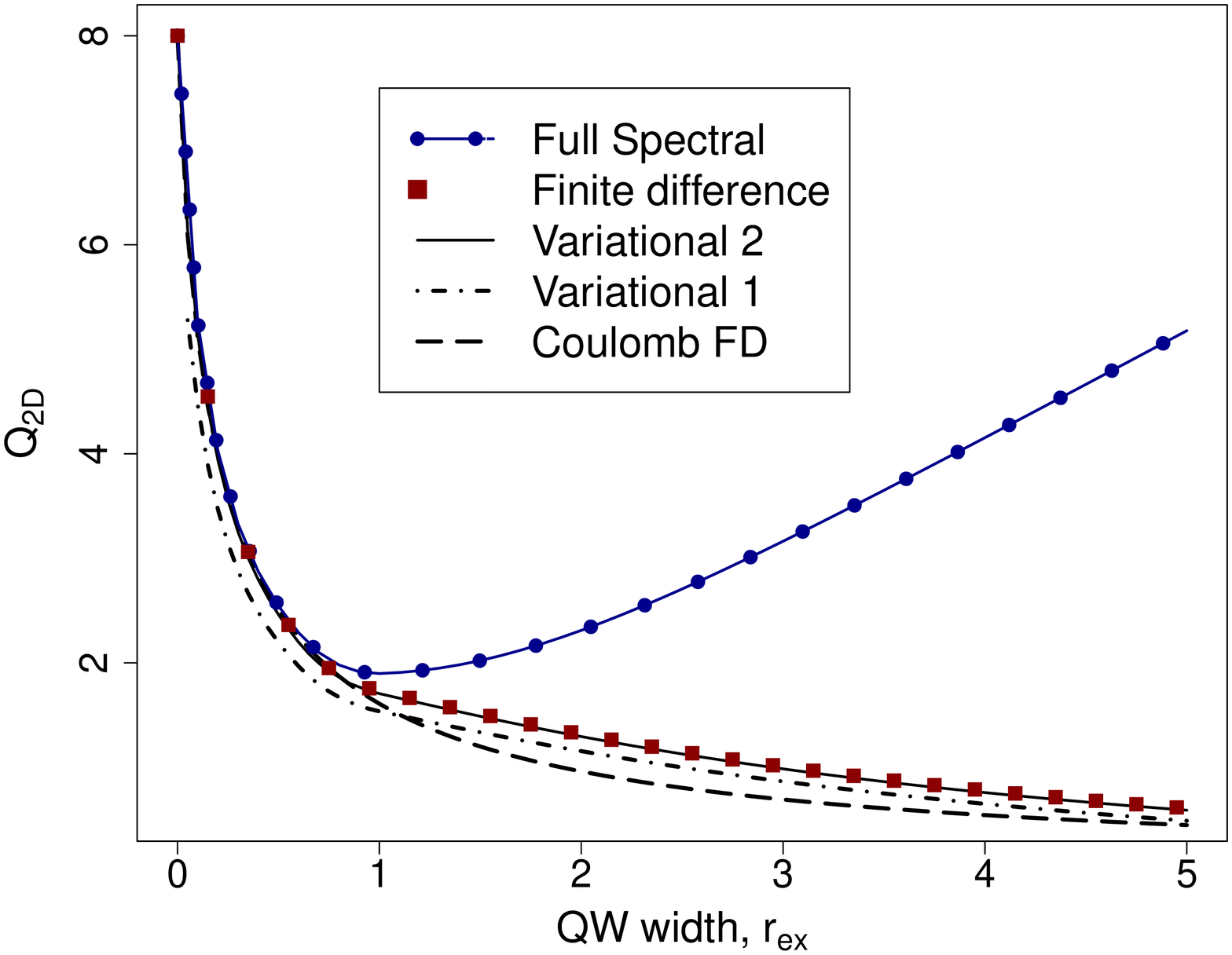}
$ \qquad $  (c)$ \qquad \qquad \qquad \qquad \qquad \qquad \qquad \qquad \qquad  $(d)

\includegraphics[width=7.4cm]{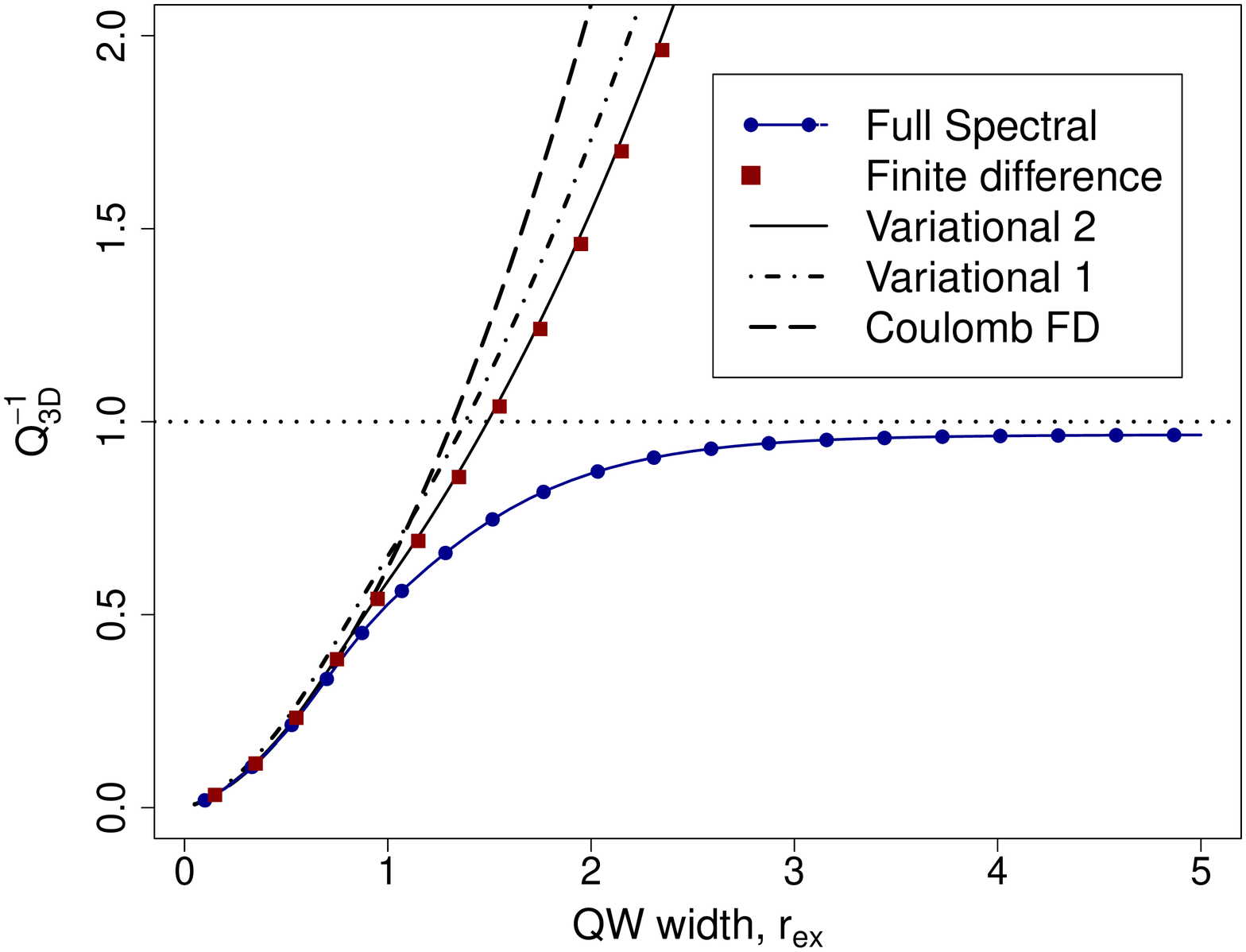}
\includegraphics[width=7.4cm]{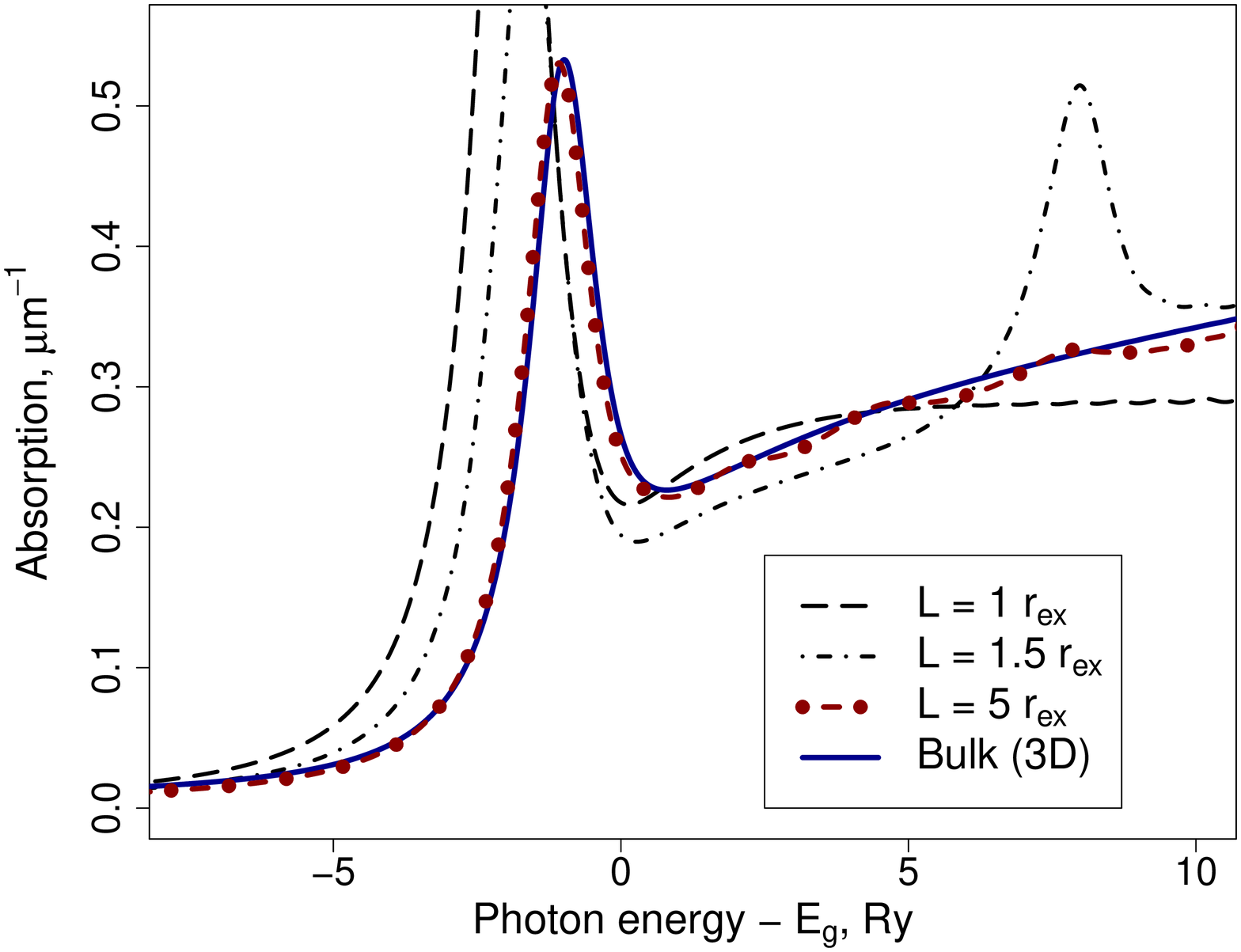}
$ \qquad $  (e)$ \qquad \qquad \qquad \qquad \qquad \qquad \qquad \qquad \qquad  $(f)

\caption{Absorption spectra, calculated with different methods for (a) $L = 0.5~ r_{ex} \approx 10~ \text{nm}$ and (b) $L = 1~ r_{ex} \approx 20 ~\text{nm}$. (c) Exciton binding energy. (d) 2D dimensionless oscillator strength. (e) Reciprocal 3D dimensionless oscillator strength, proportional to peak magnitude. (f) Absorption spectra, calculated with the spectral method for different QWs compared with the bulk (3D) case, calculated with FD.}
\end{figure}

We have calculated the absorption spectra without electric field for QW widths ranging from effectively $0$ (the 2D strong confinement regime) to $L = 5 ~ r_{ex}$, which, as we can see from Figure 3(c,e,f), gives almost identical results to the bulk 3D case.

We have used the following parameters: $R=500$ nm for all the methods, for the spectral method, $N_z=30$ and $N_r=150$; for the finite difference method (within the separation of variables approach) $\Delta r = 0.2$ nm.

As can be seen from Figure 3(a-e), there's excellent agreement for $L \leq r_{ex}$ between all the calculation methods for the Gaussian interaction potential, as well as the first absorption peak for Coulomb excitons (calculated with separation of variables approach together with the finite difference method for the interaction equation).

Simple variational ansatz with a single parameter (we denote results "Variational 1") gives a reasonably good agreement for both the energy and magnitude of the first peak, but a more elaborate three-parameter choice ("Variational 2") provides a much better accuracy, practically indistinguishable from numerically exact finite difference calculations.

When it comes to $L > r_{ex}$, the spectral solution of the full excitonic equation with no separation of variables assumption (we denote it "Full spectral") gives results qualitatively different from the approximations.

It should be noted, that piecewise fit of the Gaussian potential parameters to the Coulomb case does not change the qualitative results, only provides the necessary quantitative agreement in the range of applicability of the separation of variables approximation. Keeping the parameters $a,b$ constant, for example, fitting them to the 3D Coulomb case only, still gives the same qualitative outcome when it comes to the $L-$dependence, namely, monotonic decrease of the binding energy (Figure 3c) and monotonic decrease of the peak magnitude (Figure 3e) from pure 2D to pure 3D values. The separation of variables approach only works for $L \leq r_{ex}$ for either constant or variable interaction potential parameters.

Dimensionless quantities proportional to oscillator strength $Q_{2D}$ and $Q_{3D}$ provide a useful comparison between QWs of different width. We can see that for $L > r_{ex}$ the full spectral solution gives a physically correct picture where $Q_{2D}$ experiences linear growth (Figure 3d), while $Q_{3D}$ approaches a constant value for free bulk excitons (Figure 3e).

\subsection{Dependence on electric field strength}

\begin{figure}
\centering
\includegraphics[width=7.4cm]{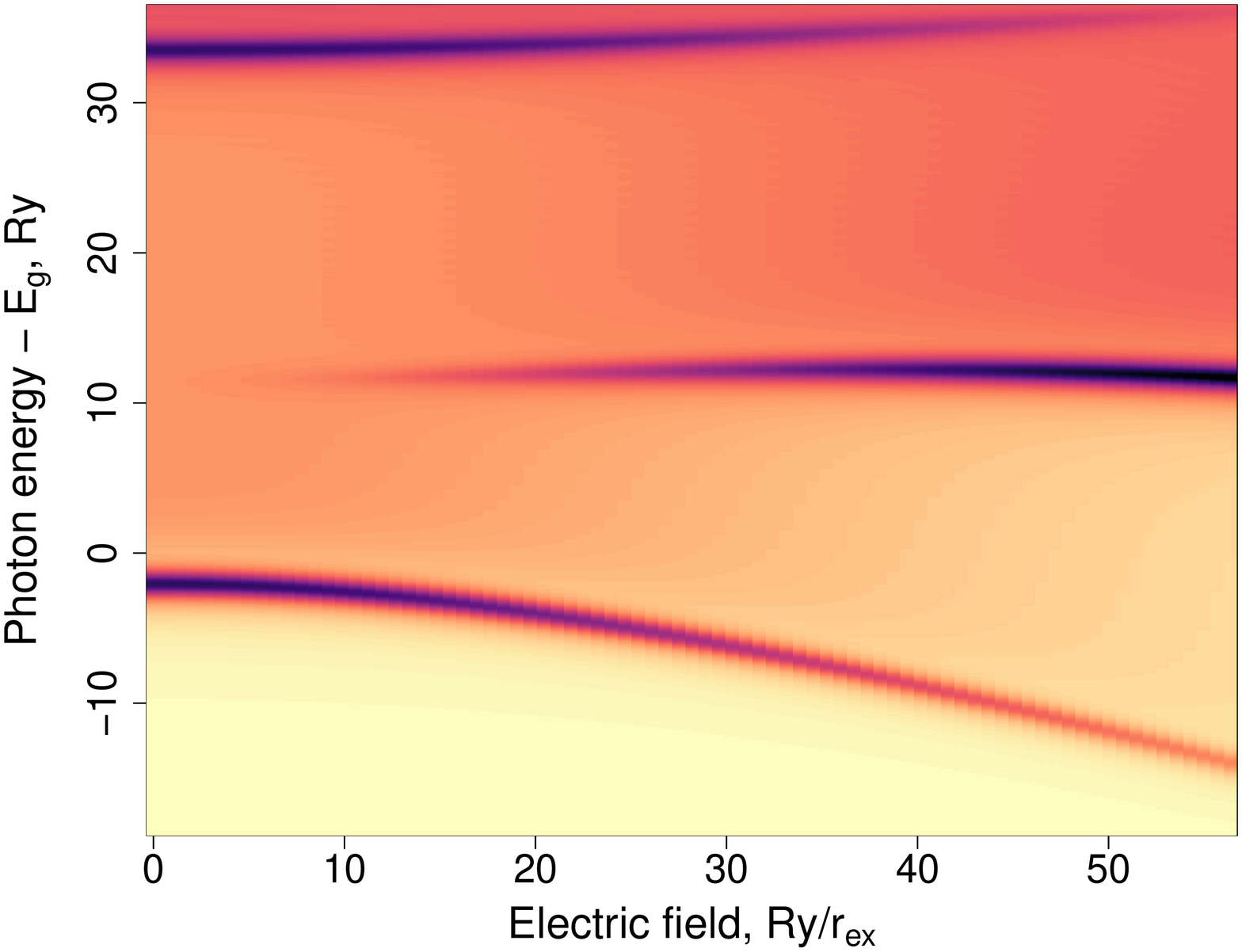}
\includegraphics[width=7.4cm]{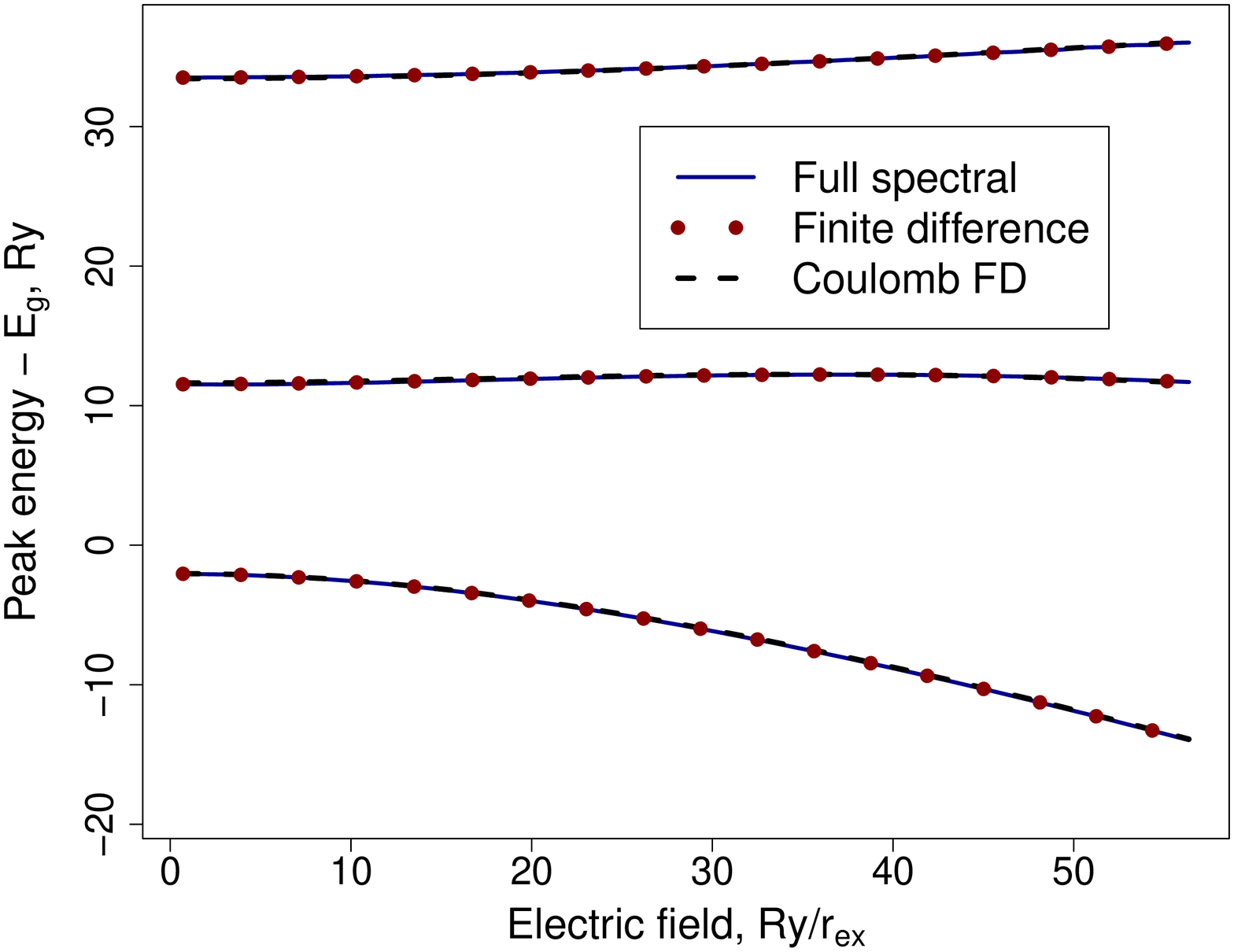}
$ \qquad $  (a)$ \qquad \qquad \qquad \qquad \qquad \qquad \qquad \qquad \qquad  $(b)

\includegraphics[width=7.4cm]{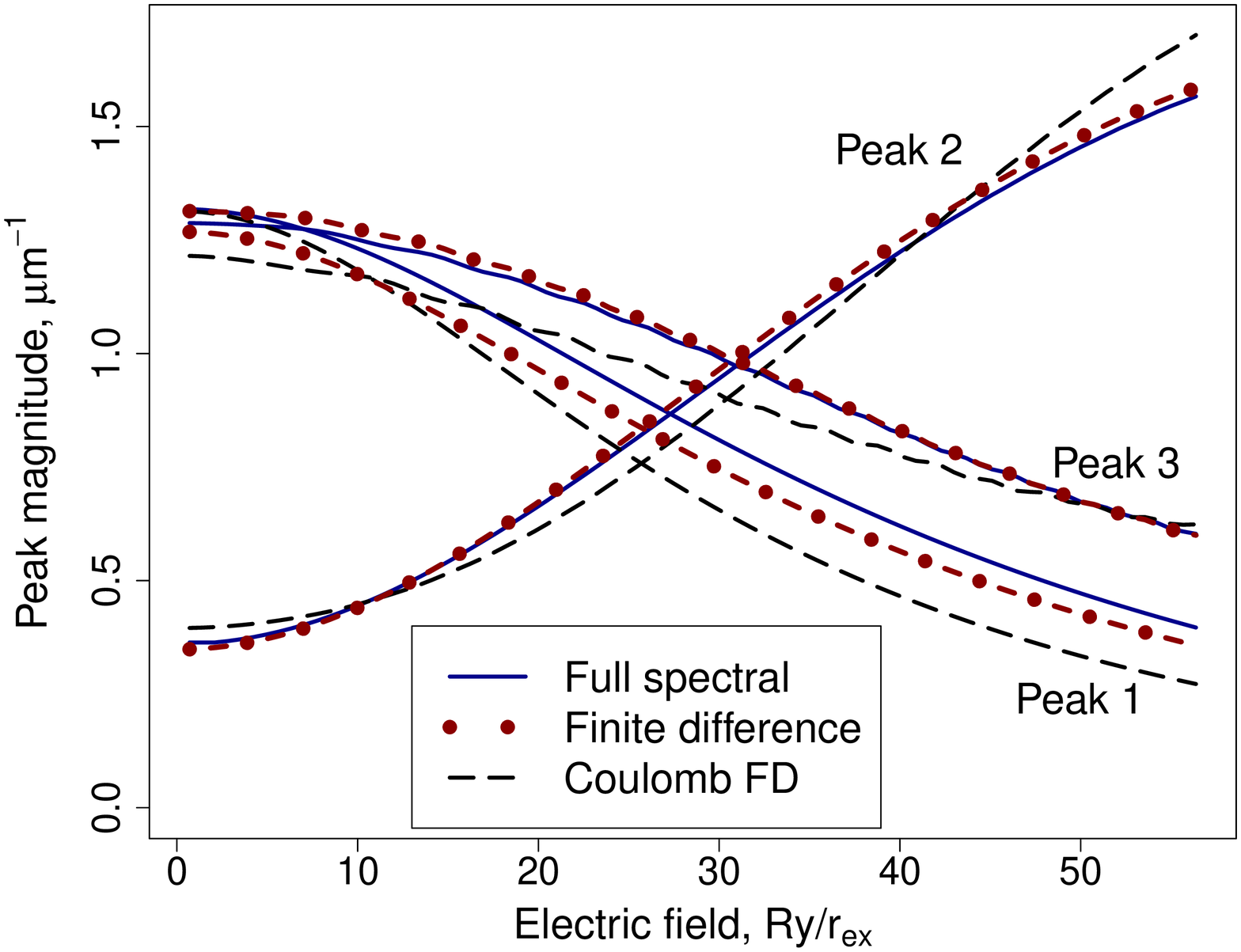}
\includegraphics[width=7.4cm]{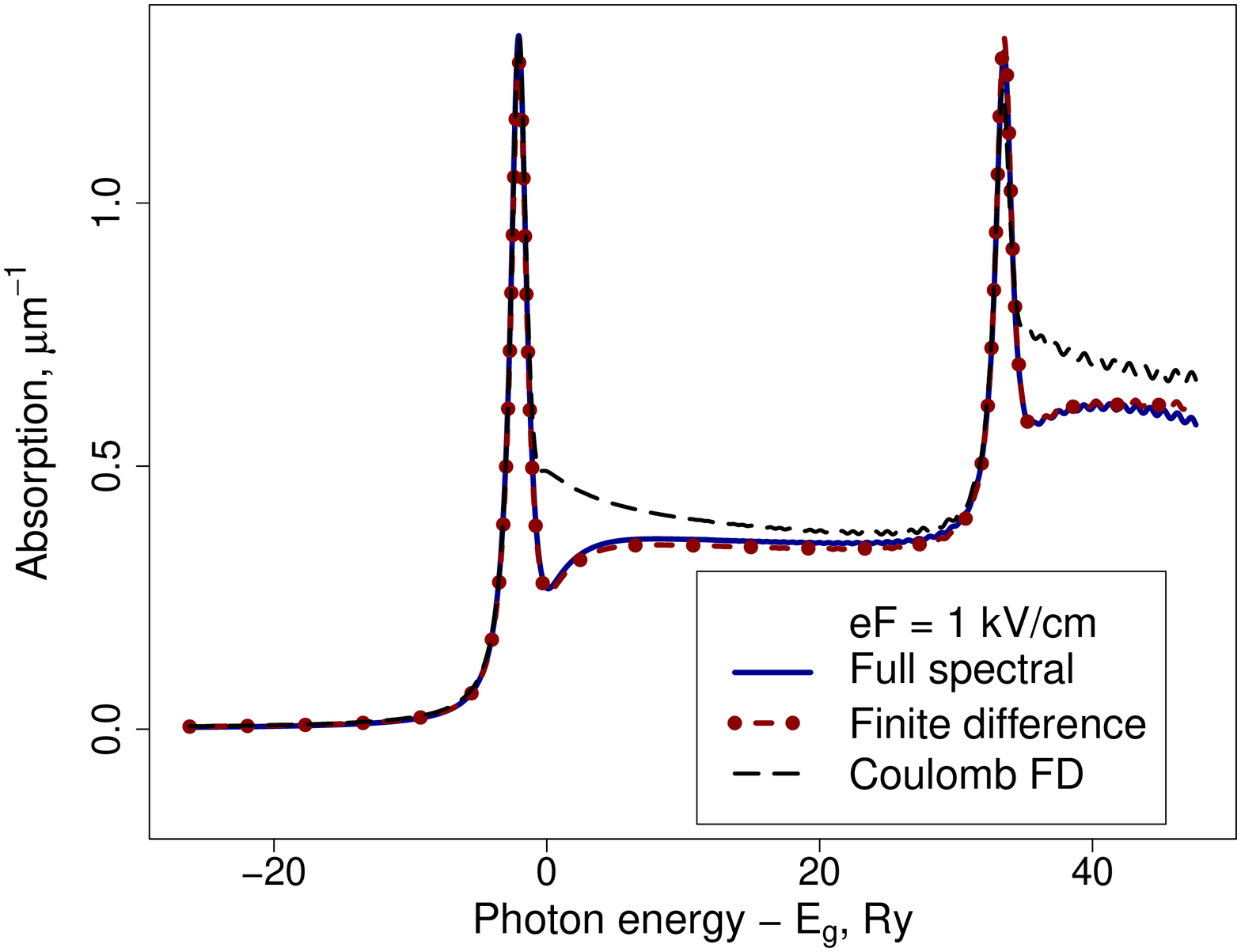}
$ \qquad $  (c)$ \qquad \qquad \qquad \qquad \qquad \qquad \qquad \qquad \qquad  $(d)

\includegraphics[width=7.4cm]{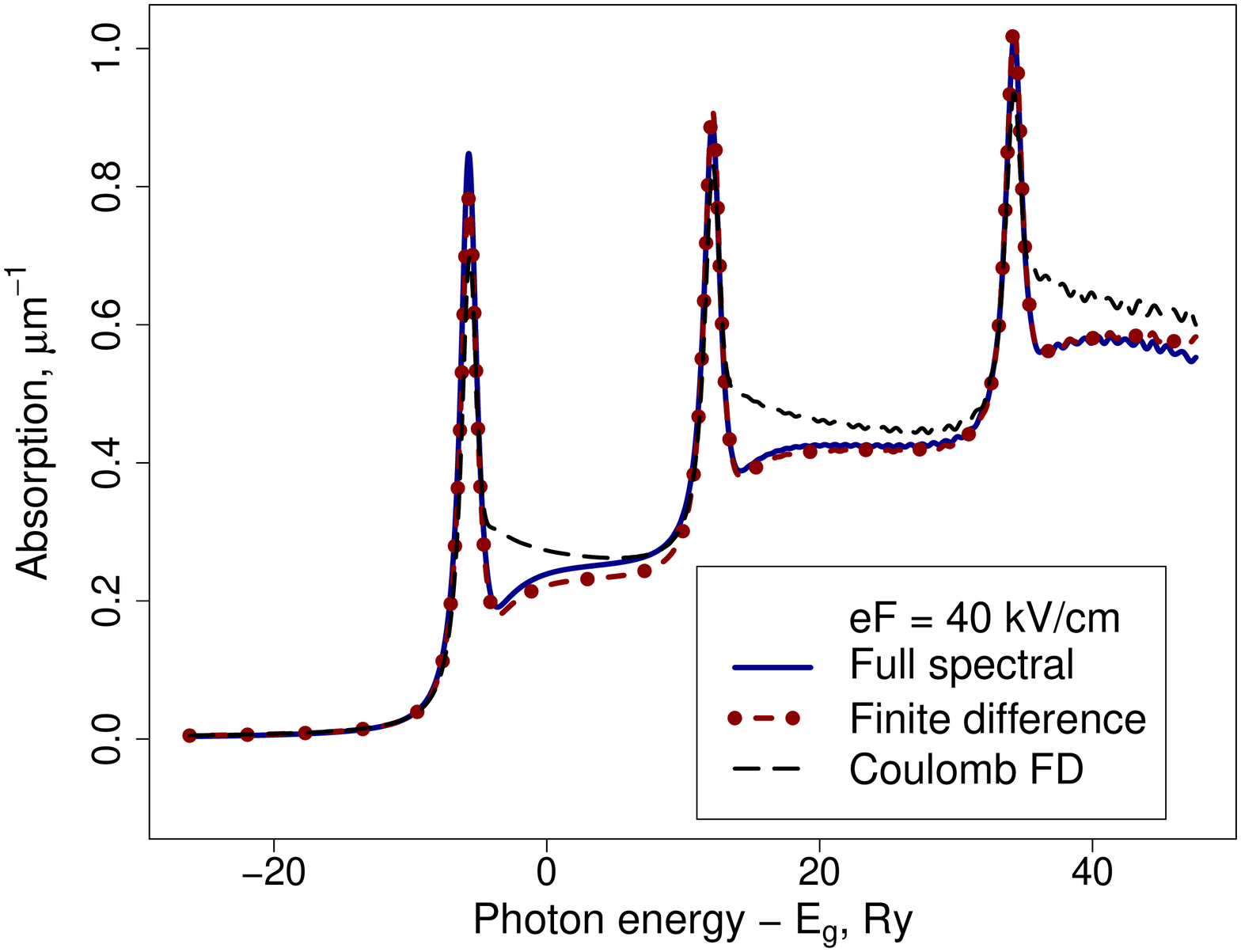}
\includegraphics[width=7.4cm]{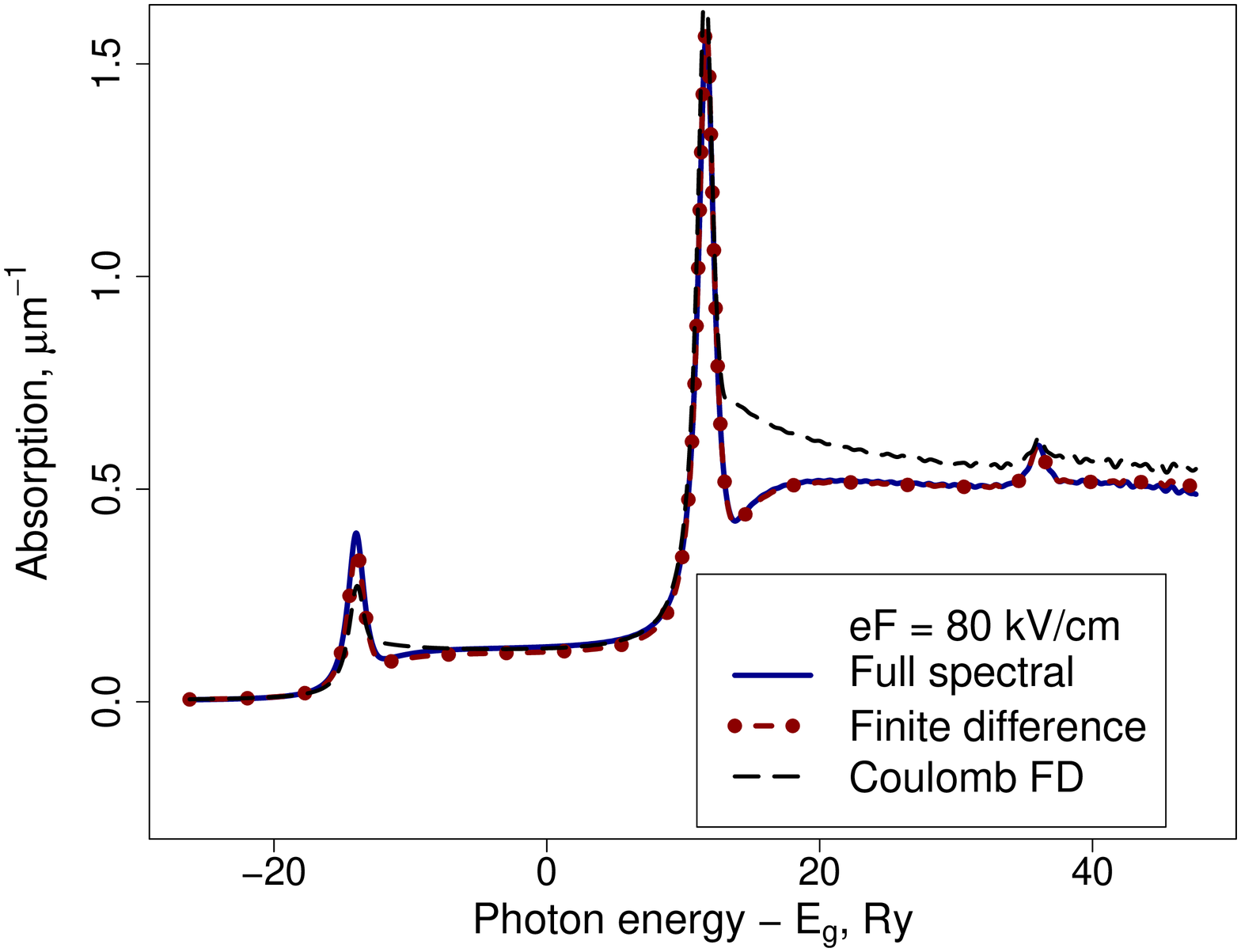}
$ \qquad $  (e)$ \qquad \qquad \qquad \qquad \qquad \qquad \qquad \qquad \qquad  $(f)

\caption{$L = 0.75 r_{ex} \approx 15 \text{nm}$: (a) Absorption spectrum dependence on the electric field. Color describes magnitude of the absorption coefficients. (b) Positions of absorption peaks. (c) Magnitudes of absorption peaks. Absorption spectra for different electric field strengths: (d) $F = 1$ kV/cm, (e) $F = 40$ kV/cm, (f) $F = 80$ kV/cm.}
\end{figure}

\begin{figure}
\centering
\includegraphics[width=7.4cm]{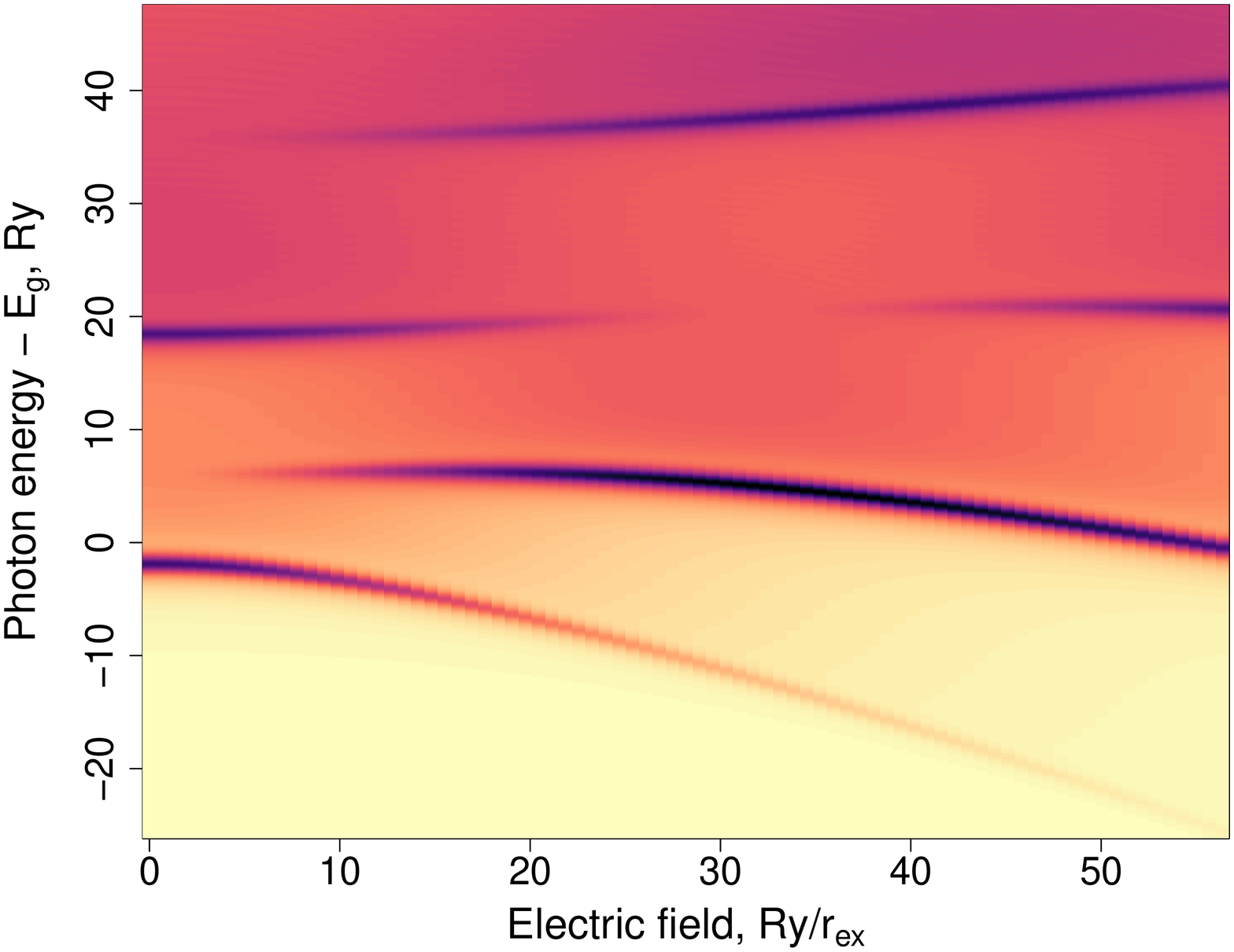}
\includegraphics[width=7.4cm]{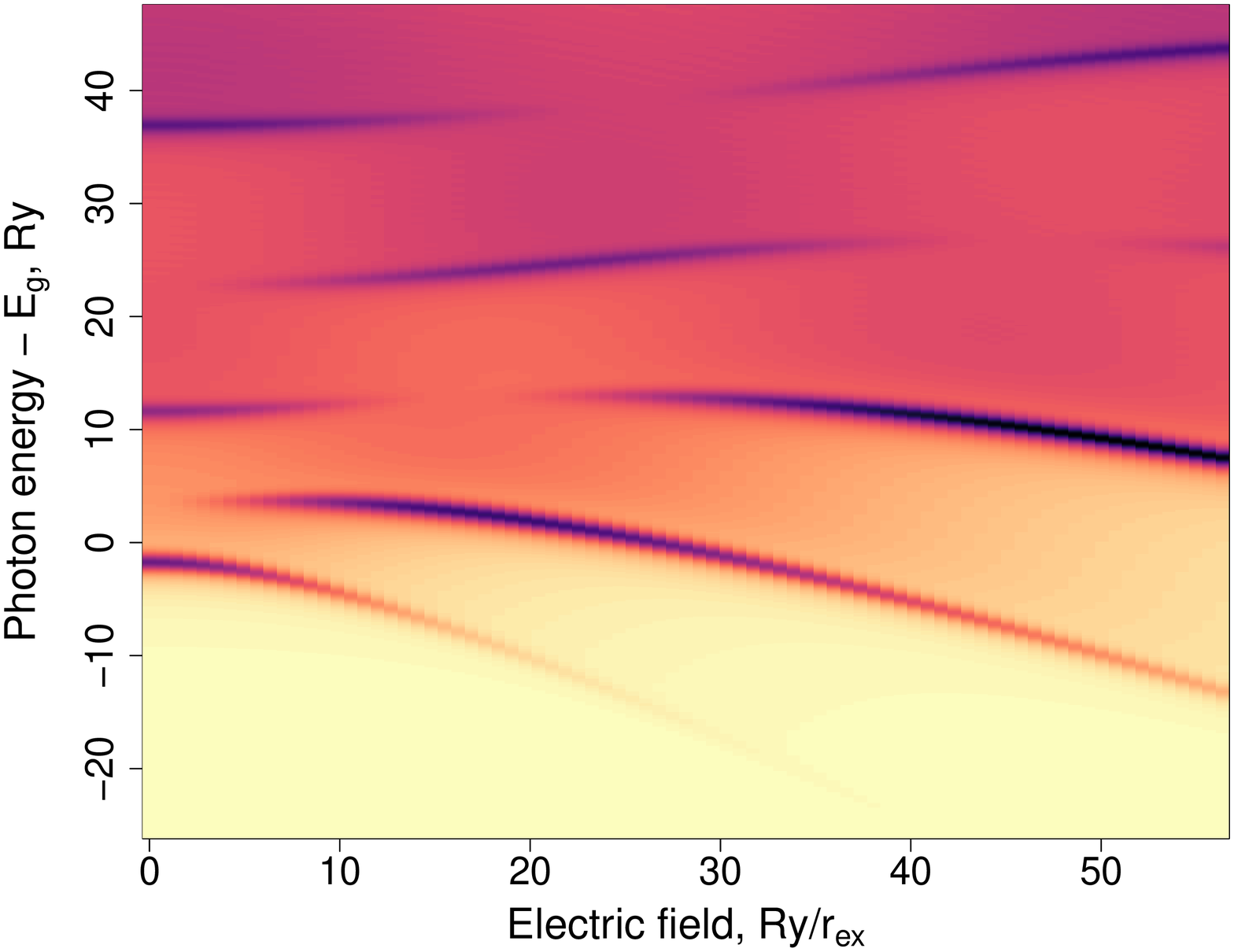}
$ \qquad $  (a)$ \qquad \qquad \qquad \qquad \qquad \qquad \qquad \qquad \qquad  $(b)

\includegraphics[width=7.4cm]{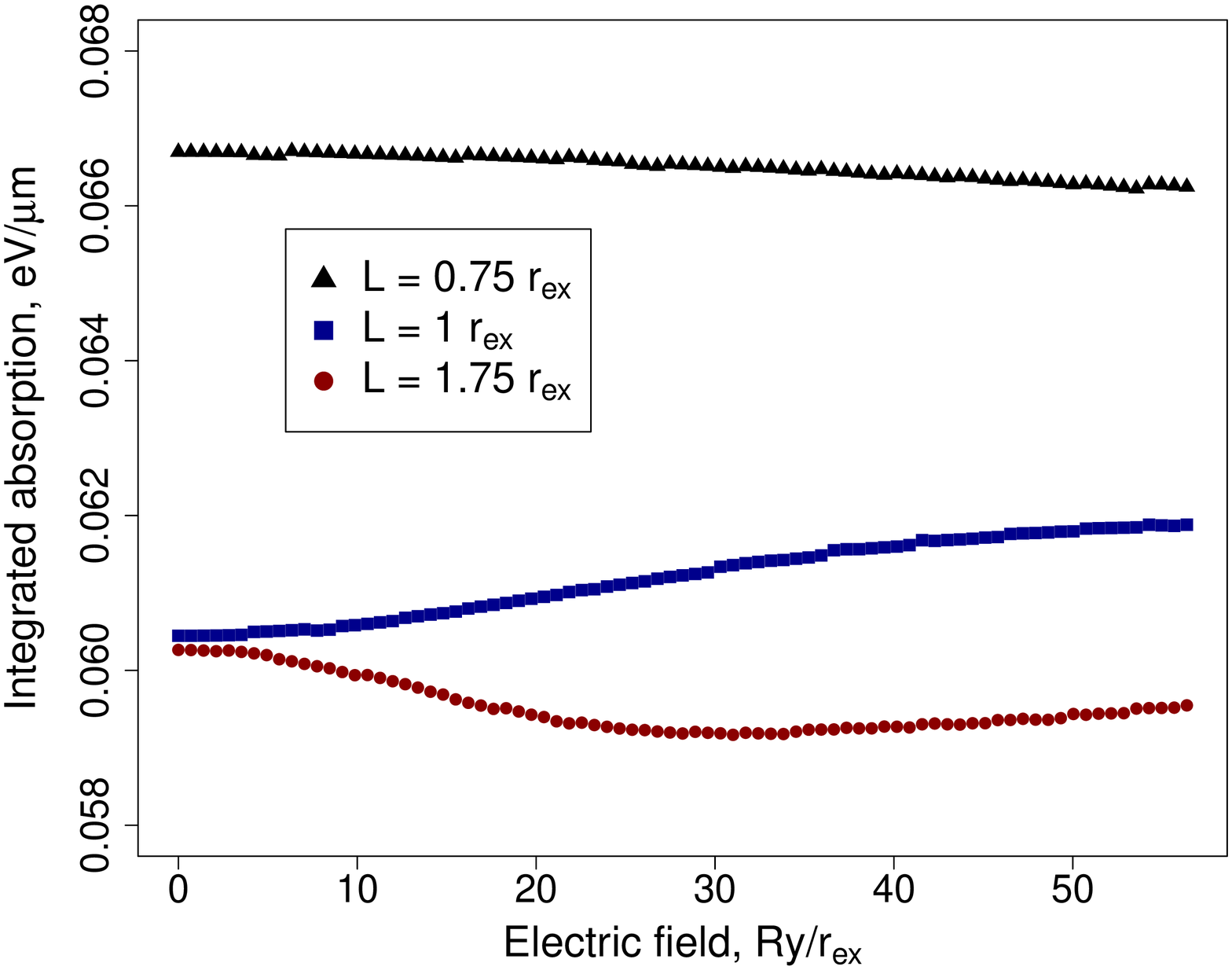}
\includegraphics[width=7.4cm]{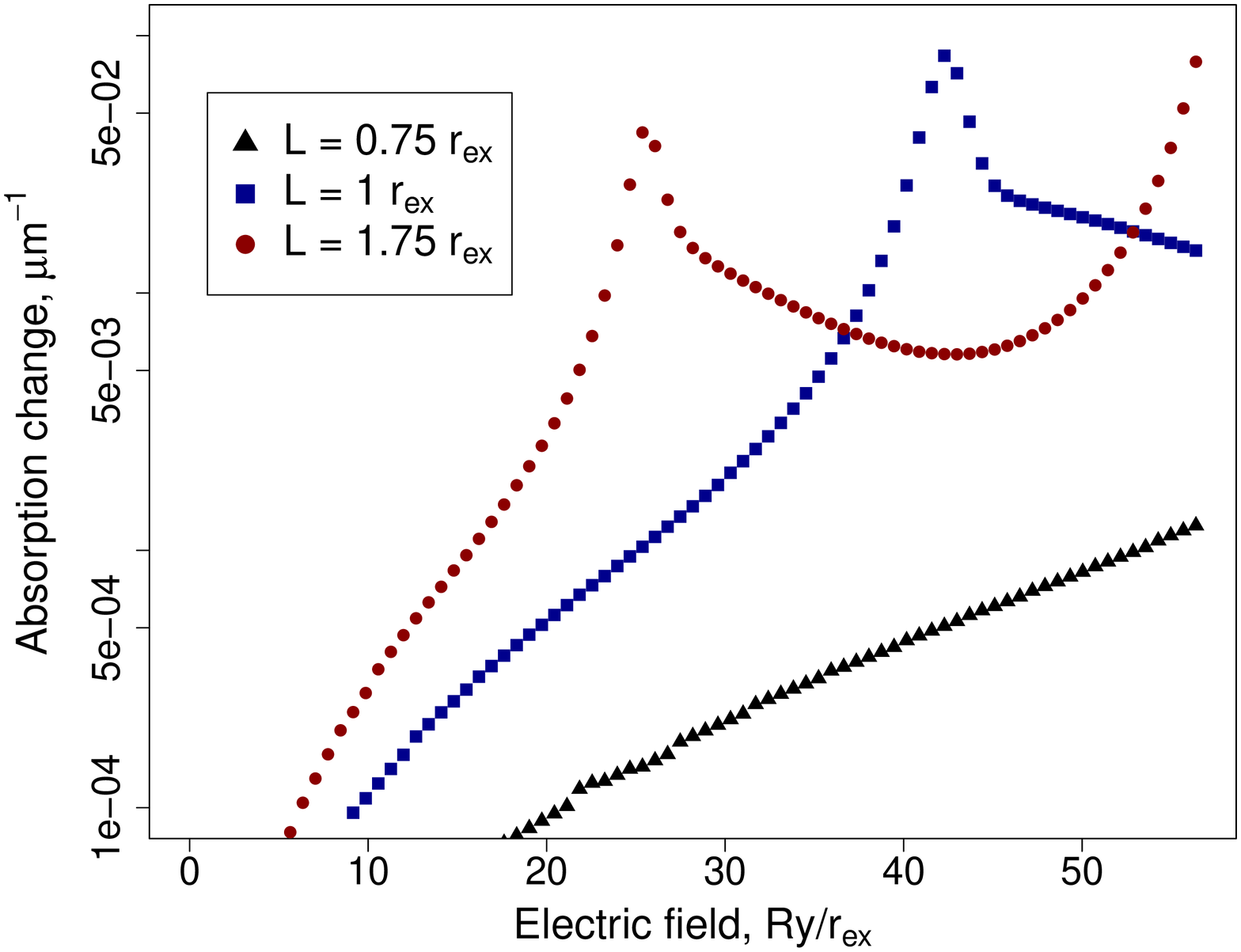}
$ \qquad $  (c)$ \qquad \qquad \qquad \qquad \qquad \qquad \qquad \qquad \qquad  $(d)

\caption{Absorption spectra dependence on the electric field. Color describes magnitude. (a) $L = 1~ r_{ex} \approx 20~ \text{nm}$, (b) $L = 1.25 ~r_{ex} \approx 25 ~\text{nm}$. (c) Dependence of integrated absorption spectrum on the electric field. (d) Electroabsorption effect at $\hbar \omega = 0.7$ eV, log scale.}
\end{figure}

Figure 4(a) shows the field dependence of the absorption spectrum for a QW with $L=0.75~r_{ex}$ which is approximately equal to $15 nm$ for the material considered in this work. The color describes the magnitude of the absorption, dark bands are the absorption peaks. Note, that we used dimensionless excitonic units for the field, while the range in absolute units is $F = [0,80]$ kV/cm.

We can observe that both magnitude and position of the peaks change with the field, and while the ground state transition is suppressed by the field, the previously forbidden second transition is, to the contrary, enhanced.

Due to the small width, all calculation methods are more or less in agreement, including the results for the Coulomb exciton (obtained, as mentioned above, using the separation of variables approach). There's a better agreement between peak positions than peak magnitudes (Figures 4(b,c)).

The main difference between the spectra for the Coulomb and Gaussian excitons is the Sommerfeld enhancement, which is much stronger for the former (Figures 4(d-f)).

Figures 5(a,b) for wider QWs provide more information about the effects of the electric field on the absorption. If we number the QW states in order of increasing energy as $\lambda=1,2,3,\ldots$, then in the absence of the electric field ($F=0$) every odd numbered state corresponds to an allowed transition, while every even numbered state - to a forbidden one. This is because in our initial problem statement we only considered the relative motion of the electron-hole pair. Which means that when $z=0$ both electron and hole have the same absolute coordinate ($z_e=z_h$). If the relative motion wavefunction has a node at this point (which is true for all the even numbered states), then absorption is forbidden.

Application of the electric field displaces both particles in the opposite directions, which leads to a deformation of the relative motion wavefunction. Depending on the number of nodes, the optical transitions are either suppressed or enhanced at small fields. For multi-node states (with $\lambda>2$) we observe non-monotonic behavior of the absorption peak magnitude. This is caused by the fact that with increasing field strength several maxima of electron and hole respective wavefunctions are displaced in the opposite directions and periodically overlap.

At very large fields, all the absorption peaks are eventually displaced in the direction of smaller energies (larger wavelengths) and their magnitude decreases until they are suppressed completely.

In addition, we have plotted field dependence of the integrated absorption spectra for three different QWs (Figure 5(c)). There's a so called "sum rule" \cite{Miller2} which states that this quantity should stay constant. In our case, it's not fulfilled exactly, but the change is only about $1-2 \%$ even at relatively large fields. The discrepancy can be easily explained by the finite range of photon energies $\hbar \omega$ we considered. Which means some exciton peaks, which remained out of range in our calculations, should affect the integrated absorption and provide a better agreement with the sum rule.

Finally, we have calculated the electroabsorption effect (the absolute change of absorption coefficient due to applied field) at $\hbar \omega=0.7 $ eV (Figure 5(d)). This energy is below the bulk bandgap when no field is applied. In this calculation, unlike all the previous ones, we have accounted for the full quantum size effect for each quantum well, by adding the following quantity to the transitions energies:

\begin{equation*}
\Delta E= \frac{\pi^2 \hbar^2}{2 m_{eh} L^2} 
\end{equation*}

This ensured that the relative positions of the absorption peaks are correct. The results (presented on a logarithmic scale) indicate initial exponential growth below the bandgap, which is stronger for larger QWs. However, due to the displacement of the ground state absorption peak, there's a critical field strength at which the absorption increase is replaced by decrease. This critical value is greater for smaller QWs, which makes them more convenient for electroabsorption modulator applications.

\section{Conclusion}
\label{sec6}

In conclusion, we have numerically investigated excitonic electroabsorption in a quantum well, using a modified interaction potential, namely a Gaussian function. The purpose of this work was to utilize the favorable properties of the latter (namely, the lack of singularity at $r=0$ and finite range) to simplify the numerical solution of Schr\"{o}dinger equation for confined excitons under external electric field.

We have developed a finite difference scheme for the radial Schr\"{o}dinger equation with either axial or central symmetry, using weighted averaging for Coulomb potential at $r=0$. The FD calculations are in excellent agreement with known analytical expressions for the absorption spectrum.

We have managed to fit the free parameters of the Gaussian potential (magnitude and effective range) to the usual Coulomb case (for the first absorption peak) for 3D and 2D excitons. We then used the separation of variables approach to obtain an approximate solution for both cases applicable to thin quantum wells ($L<r_{ex}$). This allowed us to achieve a good agreement between the Coulomb and Gaussian excitons for arbitrary QW widths. Thus, we can expect that electroabsorption modeling conducted in the present work can be used to analyze the behavior of actual excitons not only qualitatively, but in some cases quantitatively as well, especially when it comes to the absorption edge.

As for the numerical methods used in this work, the most effective way to solve the full exciton Schr\"{o}dinger equation in 2 variables proved to be the spectral expansion method. We have managed to find either exact or approximate expressions for all the matrix elements and find the absorption spectra for a wide range of QW widths. The results for very wide wells agree with bulk (3D) calculations performed by finite differences.

As a promising way to decrease the calculation time, we have also used another approach - the separation of variables. In this case, the interaction potential, which can't be separated exactly, is replaced by its average over the extra variable(s). To solve the resulting equation, one can either use a single parameter variational ansatz, a multi-parameter variational ansatz, or a finite difference scheme. We found that the first approach, while being the most simple and leading in this case to a semi-analytical solution, doesn't give the best agreement with the original two-variable equation. Meanwhile, a more advanced variational function or numerical finite difference solution provide an almost perfect agreement for $L<r_{ex}$.

Regarding the electric field dependence of the absorption spectrum, we find that for small fields the results agree well with the perturbation theory treatment, i.e. the ground state energy decreases, while all the rest of the energy levels increase. The displacement of the electron and hole wavefunctions to the opposite sides of the quantum well results in either decreasing or increasing magnitude of the absorption peaks, which is related to the symmetry and relative position of the wavefunction extrema. Which is why for excited states with several such extrema we observe non-monotone field dependence of both the magnitude and transition energy.

As for electroabsorption below the edge, the magnitude of the effect varies with QW width, and for certain fields we find that quantum wells of intermediate thickness perform better than those that are either too thin or too thick.

The main conclusions of this work are in good agreement with previously published results on Coulomb excitons, which means that the numerically stable Gaussian interaction potential may be useful for modeling more complex phenomena as well, especially with many interacting particles or fast-changing fields.

\section*{Acknowledgments}
The research was carried out within the state assignment of FASO of Russia \textit{N}\textsuperscript{\underline{o}} 0723-2020-0037.



\end{document}